\documentclass[3p]{elsarticle}
\UseRawInputEncoding
\usepackage{lineno,hyperref}
\usepackage{graphicx}
\usepackage{amsmath}
\usepackage{color}
\usepackage{mathtools}
\usepackage[normalem]{ulem}
\usepackage{rotating}
\usepackage{lscape}

\journal{economies@mdpi}
\begin{document}

\begin{frontmatter}

\title{Exploring Distributions of House Prices and House Price Indices}

\author[mymainaddress]{Jiong Liu}
\author[mymainaddress]{Hamed Farahani}
\author[mymainaddress]{R. A. Serota\fnref{myfootnote}}
\fntext[myfootnote]{serota@ucmail.uc.edu}

\address[mymainaddress]{Department of Physics, University of Cincinnati, Cincinnati, Ohio 45221-0011}

\begin{abstract}
We use house prices (HP) and house price indices (HPI) as a proxy to income distribution. Specifically, we analyze sale prices in the 1970-2010 window of over 116,000 single-family homes in Hamilton County, Ohio, including Cincinnati metro area of about 2.2 million people. We also analyze HPI, published by Federal Housing Finance Agency (FHFA), for nearly 18,000 US ZIP codes that cover a period of over 40 years starting in 1980's. If HP can be viewed as a first derivative of income, HPI can be viewed as its second derivative. We use generalized beta (GB) family of functions to fit distributions of HP and HPI since GB naturally arises from the models of economic exchange described by stochastic differential equations. Our main finding is that HP and multi-year HPI exhibit a negative Dragon King (nDK) behavior, wherein power-law distribution tail gives way to an abrupt decay to a finite upper limit value, which is similar to our recent findings for realized volatility of S\&P500 index in the US stock market. This type of tail behavior is best fitted by a modified GB (mGB) distribution. Tails of single-year HPI appear to show more consistency with power-law behavior, which is better described by a GB Prime (GB2) distribution. We supplement full distribution fits by mGB and GB2 with direct linear fits (LF) of the tails. Our numerical procedure relies on evaluation of confidence intervals (CI) of the fits, as well as of p-values that give the likelihood that data come from the fitted distributions.
\end{abstract}


\end{frontmatter}

\section{Introduction\label{Introduction}}

Income distributions have a long history of being modeled by the Generalized Beta (GB) family of distributions. A comprehensive summary can be found in \cite{chotikapanich2008modelling}, and specifically in \cite{mcdonald2008modelling}, while more recent works include \cite{chotikapanich2018using} and \cite{dashti2020stochastic}. The motivation for this study was to glean insight into whether income distributions exhibit power-law (``fat") tails or display outliers such as Dragon Kings (DK) or negative Dragon Kings (nDK) \cite{sornette2012dragon,wheatley2015multiple,liu2023rethinking,liu2023dragon}. Given that identifying fat tails and outliers requires vary large datasets, available coarse-grained income data is not sufficient for this purpose. In this regard, house prices (HP) may serve as a proxy to -- or ``derivative" of -- incomes. 

Previously, we considered sale prices of houses in Hamilton County, Ohio from 1970 to 2010 in the context of analysis of inequality indices \cite{dashti2020stochastic}. Here we pare down that data to include only single-family homes, which better aligns to incomes, as well as to house price indices (HPI), which are also based on single-family HP and are constructed using repeat-sale methodology \cite{baily1963regression}, \cite{case1987prices}, \cite{case1989efficiency}, \cite{calhoun1996house}, \cite{bogin2016local}. Specifically, we utilize FHFA HPI for nearly 18,000 US ZIP codes over a period of 40 years, starting in 1980's \cite{bogin2016local}. HPI can be viewed as a proxy to HP -- or a "second derivative" of income distribution -- in this case between different ZIP codes.

Our approach to analyzing HP and HPI distributions mirrors the approach we used for analysis of the distributions of historical realized volatility \cite{liu2023dragon}. Namely, we fitted the entire distribution with modified Generalized Beta (mGB) and Generalized Beta Prime (GB2) distributions, as well as performed a linear fit (LF) of the tails. For each fit, we evaluated confidence interval (CI) as well as conducted a U-test \cite{liu2023dragon},\cite{pisarenko2012robust}, which yields p-values that indicate the likelihood of the data coming from the fitted distribution.

\newpage
Since fat tails are scale free, they are naturally analyzed on a log-log scale, where complementary cumulative distribution function (CCDF) ought to be a straight line with negative slope. LF of CCDF, however, must be supplemented by a statistical test to determine the likelihood that  the points in the tail conform to linearity or whether outliers may be present \cite{wheatley2015multiple}. In this case we are concerned with possibility of observing DK, where the ends of the tails shoot upward from LF, or nDK, where tails' ends shoot down. 

Towards this end, we performed a U-test \cite{pisarenko2012robust} but we did not limit it to LF alone. We also aimed at describing the entire distribution, not just tails, and thus performed a U-test on mGB and GB2, which we employed for this purpose. GB2 was used for fitting since it is the most flexible distribution with fat tails and mGB, while exhibiting similar tails over a wide range of variable, abruptly terminates at finite value, which seems appropriate to describe nDK behavior \cite{liu2023rethinking}. Of particular importance is the fact that both mGB and GB2 arise as steady-state (stationary) distributions of stochastic differential equations that mean to serve as models of economic exchange \cite{dashti2020stochastic,bouchaud2000wealth,ma2013distribution}. 

This paper is organized as follows. In Section \ref{GBDF} we present the analytical form of mGB and GB2 distributions and discuss the limiting behaviors of both. In Section \ref{HPHPIfit} we fit HP and HPI with mGB and GB 2, as well as the fit tails directly using LF. For each test we conduct a U-test, for which a null hypothesis is formulated, and plot p-values which reflect on the goodness of fit, as well as whether DK or nDK behavior may be present. We conclude in Section \ref{Discussion} with discussion of our results.

\section{Generalized Beta Distribution Function \label{GBDF}}

A detailed discussion of the Modified Generalized Beta (mGB) distribution function, used here to fit the distributions of HP and HPI, as well as of the Generalized Beta (GB) family of distributions in general, can be found in \cite{liu2023rethinking, liu2023dragon}. A generalization of the traditional GB distribution \cite{mcdonald1995generalization} can be written (in a slightly modified form relative to that of \cite{liu2023rethinking}) as follows \cite{liu2023dragon}:
\begin{equation}
f_{GB}(x;\alpha,\beta _1,\beta _2, p,q)= \frac{\alpha \left(1+\left(\frac{\beta _2}{\beta _1}\right)^{\alpha }\right)^p \left(\frac{x}{\beta _2}\right)^{\alpha  p-1} \left(1+\left(\frac{x}{\beta _2}\right)^{\alpha}\right)^{-p-q} \left(1-\left(\frac{x}{\beta _1}\right)^{\alpha }\right)^{q-1}}{\beta _2 B(p,q)},
\label{GBPDF2}
\end{equation}
where $\beta _1$ and $\beta _2$ are scale parameters and $\alpha$, $p$ and $q$ are shape parameters, all positive, $B(p,q)$ is the beta function and $x \leq \beta _1$. Although it has a concise and transparent form, it does not come out as a solution of a stochastic differential equation (SDE) \cite{hertzler2003classical}, the latter being desirable for the purpose of modeling behavior of quantities, such as stochastic volatility, important for understanding of implied and realized volatility in equity markets \cite{dashti2021combined}, and income distribution, resulting from economic exchange \cite{dashti2020stochastic,bouchaud2000wealth,ma2013distribution}.
 
The probability density function (PDF) of mGB, which comes out as a solution of an SDE (with minor caveats explained in \cite{liu2023rethinking}) and which is used here to model HP and HPI distributions, can be written as 
\begin{equation}
f_{mGB}(x;\alpha,\beta _1,\beta _2, p,q)=\frac{\alpha(p+q)\left(1+\left(\frac{\beta _2}{\beta _1}\right)^\alpha\right)^{p+1}  \left(\frac{x}{\beta _2}\right)^{\alpha p-1}  \left(1+\left(\frac{x}{\beta _2}\right)^\alpha\right)^{-p-q-1} \left(1-\left(\frac{x}{\beta _1}\right)^\alpha\right)^{q-1}}{\beta _2 B(p,q)\left(q+\left(\frac{\beta _2}{\beta _1}\right)^\alpha (p+q) \right) },
\label{mGBPDF}
\end{equation}
where $\beta _1$ and $\beta _2$ are scale parameters and $\alpha$, $p$ and $q$ are shape parameters, all positive, $B(p,q)$ is the beta function and $x \leq \beta _1$. 
The cumulative distribution function (CDF) and complementary CDF (CCDF) of mGB are given respectively by 
\begin{equation}
\resizebox{1.0\hsize}{!}{$
F_{mGB}(x;\alpha,\beta _1,\beta _2, p,q)=I\left({\frac{\left(\frac{x}{\beta _1}\right)^\alpha+\left(\frac{x}{\beta _2}\right)^\alpha}{1+\left(\frac{x}{\beta _2}\right)^\alpha}};p,q\right) + \frac{1}{B(p,q) \left(q+\left(\frac{\beta _2}{\beta _1}\right)^{\alpha } (p+q)\right)} \left(\frac{1-\left(\frac{x}{\beta _1}\right)^\alpha}{1+\left(\frac{x}{\beta _2}\right)^\alpha}\right)^q\left(\frac{\left(1+\left(\frac{\beta _2}{\beta _1}\right)^\alpha\right) \left(\frac{x}{\beta _2}\right)^\alpha}{\left(1+\left(\frac{x}{\beta _2}\right)^\alpha\right)}\right)^p,
$}
\label{mGBCDF}
\end{equation}
and
\begin{equation}
\resizebox{1.0\hsize}{!}{$
1-F_{mGB}(x;\alpha,\beta _1,\beta _2, p,q)=I\left({\frac{1-\left(\frac{x}{\beta _1}\right)^{\alpha }}{1+\left(\frac{x}{\beta _2}\right)^{\alpha }}};q,p\right) -\frac{1}{B(p,q) \left(q+\left(\frac{\beta _2}{\beta _1}\right)^{\alpha } (p+q)\right)} \left(\frac{1-\left(\frac{x}{\beta _1}\right)^\alpha}{1+\left(\frac{x}{\beta _2}\right)^\alpha}\right)^q\left(\frac{\left(1+\left(\frac{\beta _2}{\beta _1}\right)^\alpha\right) \left(\frac{x}{\beta _2}\right)^\alpha}{\left(1+\left(\frac{x}{\beta _2}\right)^\alpha\right)}\right)^p,
$}
\label{mGBCCDF}
\end{equation}
where the first term in (\ref{mGBCDF}) and (\ref{mGBCCDF}) represent, respectively, CDF and CCDF of GB (whose PDF is given by(\ref{GBPDF2})), while $I(y;p,q)=B(y;p,q)/B(p,q)$ and $B(y;p,q)$ are, respectively, the regularized and incomplete beta functions \cite{nist2022digital}.

In what follows, we will be specifically interested in the $\beta_2\ll\beta_1$ circumstance since for $\beta_2 \ll x \ll \beta_1$ GB and mGB exhibit a power-law dependence,
\begin{equation}
f_{mGB} \propto \left(\frac{x}{\beta _2}\right)^{-\alpha (q+1) - 1}, \hspace{.3cm} 1-F_{mGB} \propto \left(\frac{x}{\beta _2}\right)^{-\alpha (q+1)};
\hspace{.5cm} f_{GB} \propto \left(\frac{x}{\beta _2}\right)^{-\alpha q - 1}, \hspace{.3cm} 1-F_{GB} \propto \left(\frac{x}{\beta _2}\right)^{-\alpha q}.  
\label{GB2tail}
\end{equation}
In the limit of $\beta_1 \rightarrow \infty$, mGB and GB become, respectively, mGB2 and GB2 (the latter also known as Generalized Beta Prime) and are given by \cite{liu2023rethinking}
\begin{equation}
f_{mGB2}(x; \alpha, \beta_2, p,q)=\frac{\alpha (p+q) \left(\frac{x}{\beta_2}\right)^{\alpha p -1} \left(1+\left({\frac{x}{\beta_2}}\right)^{\alpha}\right)^{-p-q-1}}{q\beta_2 B(p,q)},
\label{mGB2PDF}
\end{equation}
and
\begin{equation}
f_{GB2}(x;\alpha,\beta _1,\beta _2, p,q)=\frac{\alpha\left(\frac{x}{\beta _2}\right)^{\alpha  p-1} \left(1+\left(\frac{x}{\beta _2}\right)^{\alpha}\right)^{-p-q}}{\beta _2 B(p,q)}
\label{GB2PDF}.
\end{equation}
Unlike mGB and GB, for whom the power-law dependences in (\ref{GB2tail}) eventually terminate at $\beta_1$, mGB2 and GB2 will sustain these power-law dependences indefinitely.

Below, we will use (\ref{mGBCCDF}) to fit CCDF of distributions of HP and HPI. As explained in  \cite{liu2023rethinking}, mGB2 and GB2 are equivalent since $q$ and $p$ are independently defined at this level of GB family of distributions, and $q$ can be shifted by unity in the definition of mGB2/GB2. Consequently, we choose a more familiar CCDF of GB2 
\begin{equation}
1-F_{GB2}(x;\alpha,\beta _1,\beta _2, p,q)=I\left({\frac{1}{1+\left(\frac{x}{\beta _2}\right)^{\alpha }}};q,p\right)
\label{GB2CCDF}
\end{equation}
to fit CCDF of the HP and HPI data. Insofar as the main difference between mGB and GB is concerned, it is their behavior near $\beta_1$ in the present context \cite{liu2023rethinking}. Namely,
\begin{equation}
1-F_{GB} \approx \frac{1}{qB(p,q)} \left(\frac{1-\left(\frac{x}{\beta _1}\right)^\alpha}{1+\left(\frac{x}{\beta _2}\right)^\alpha}\right)^q \approx \frac{1}{qB(p,q)} \left(\frac{1-\left(\frac{x}{\beta _1}\right)^\alpha}{1+\left(\frac{\beta _1}{\beta _2}\right)^\alpha}\right)^q,
\label{GBCDFbeta1}
\end{equation}
and
\begin{equation}
1-F_{mGB} \approx \frac{1+\frac{p}{q}}{qB(p,q)} \left(\frac{1-\left(\frac{x}{\beta _1}\right)^\alpha}{1+\left(\frac{x}{\beta _2}\right)^\alpha}\right)^q \left(\frac{\beta_2}{\beta_1} \right)^{\alpha} \approx \frac{1+\frac{p}{q}}{qB(p,q)} \left(\frac{1-\left(\frac{x}{\beta _1}\right)^\alpha}{1+\left(\frac{\beta _1}{\beta _2}\right)^\alpha}\right)^q \left(\frac{\beta_2}{\beta_1} \right)^{\alpha},
\label{mGBCDFbeta1}
\end{equation}
that is $1-F_{mGB}$ drops off to zero ($F_{mGB}$ saturates to unity) faster than $1-F_{GB}$ due to the factor $\left(\frac{\beta _2}{\beta _1}\right)^\alpha$. This feature accounts for a better fit via mGB versus GB, which may be due to the fact that mGB emerges from a physically motivated stochastic model \cite{liu2023rethinking}.

\section{Fitting Distribution of House Prices and House Price Indices\label{HPHPIfit}}
In this section we perform Bayesian fitting of the entire distributions of HP and HPI using mGB, as per (\ref{mGBPDF})-(\ref{mGBCCDF}), and GB2, as per (\ref{GB2PDF}) and (\ref{GB2CCDF}). We also perform LF of the tails. Confidence intervals (CI) for the fits are evaluated via inversion of the binomial distribution \cite{janczura2012black}; $p$-values are evaluated in the framework of the U-test, which is based on order statistics \cite{pisarenko2012robust}, using the following formula:
\begin{equation}
p(x_{k,m})=1-B\left(F(x_{k,m}); k, m-k+1\right),
\label{p-value}
\end{equation}
where $x_{k,m}$ is the $k$'s member of numbers between $1$ and $m$ ordered by increasing  magnitude (HP and HPI in this case), and $F(x_{k,m})$ is the assumed CDF (mGB, GB2 and LF here). p-values are evaluated in order to test the null hypothesis \emph{$H_0$: all observations of the sample are generated by the same fitting distribution}. The p-value (\ref{p-value}) is defined as a probability of exceeding the observed value $x_{k,m}$ under the null hypothesis. If among the p-values there are some small values -- $\le0.05$ here -- then those  observed values are identified as DK with probability $1-p$. Conversely, large p-values -- $\ge0.95$ here -- are identified as nDK with the probability $p$ \cite{pisarenko2012robust}.
\subsection{Distribution of House Prices \label{HPfit}}
In this section we fit the distribution of sale prices of single-family homes in Hamilton County, Ohio between 1970 and 2010. This dataset contains 116207 entries and sale prices are converted to constant dollars to eliminate the effect of inflation. Table \ref{paramsHP} gives all estimated parameters of mGB and GB2 distributions and slopes of two linear fits: LF-1 is performed by visually excluding possible outliers \cite{pisarenko2012robust} and LF-2 by excluding HP whose sale price was above 90\% of the top HP. %Kolmogorov-Smirnov statistic of the fits are given to compare goodness of fit between mGB and GB2 and  otherwise as for guidance only since they are obtained for fits with estimated parameters.

\begin{table}[!htb]
\caption{All estimated parameters of mGB and GB2 fits of HP distribution and slopes of GB2 and of LFs}
\label{paramsHP}
\centering
\begin{tabular}{c c c c c c c c c}
 Fit&$\alpha$ & $\beta_1$&$\beta_2$&p&q& Slope of CCDF&\\%KS&\\ 
\hline
mGB&2.3717 & 4209.7557&187.7393&0.8597&0.3891&&\\%0.0366& \\   
\hline
GB2&2.8732 & &178.0461&0.6692 &1.0289&-3.9562&\\%0.0345& \\  
\hline
LF-1& & && &&-3.1880&& \\  
\hline
LF-2& & && &&-3.0796&& \\ 
\end{tabular}
\end{table}

\begin{figure}[htbp!]
	\centering
		\includegraphics[width = .77 \textwidth]{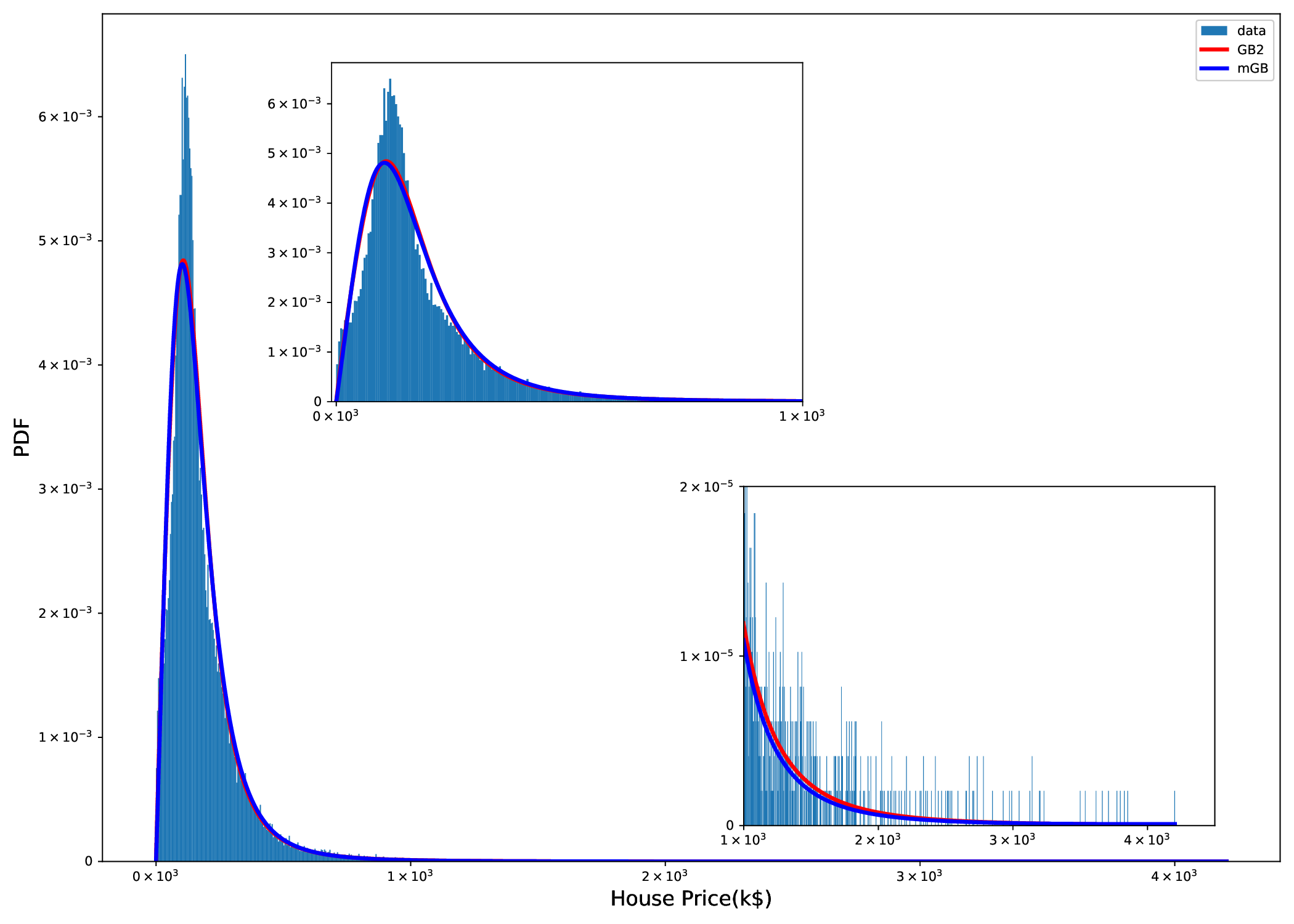}
	\caption{HP PDF: mGB and GB2 fits}
	\label{PDFHP}
\end{figure}

Fig. \ref{PDFHP} shows probability density function (PDF) of HP with mGB and GB2 fits. Clearly neither fit does a good job for typical values HP. 

\begin{figure}[htbp!]
	\centering
		\includegraphics[width = .77 \textwidth]{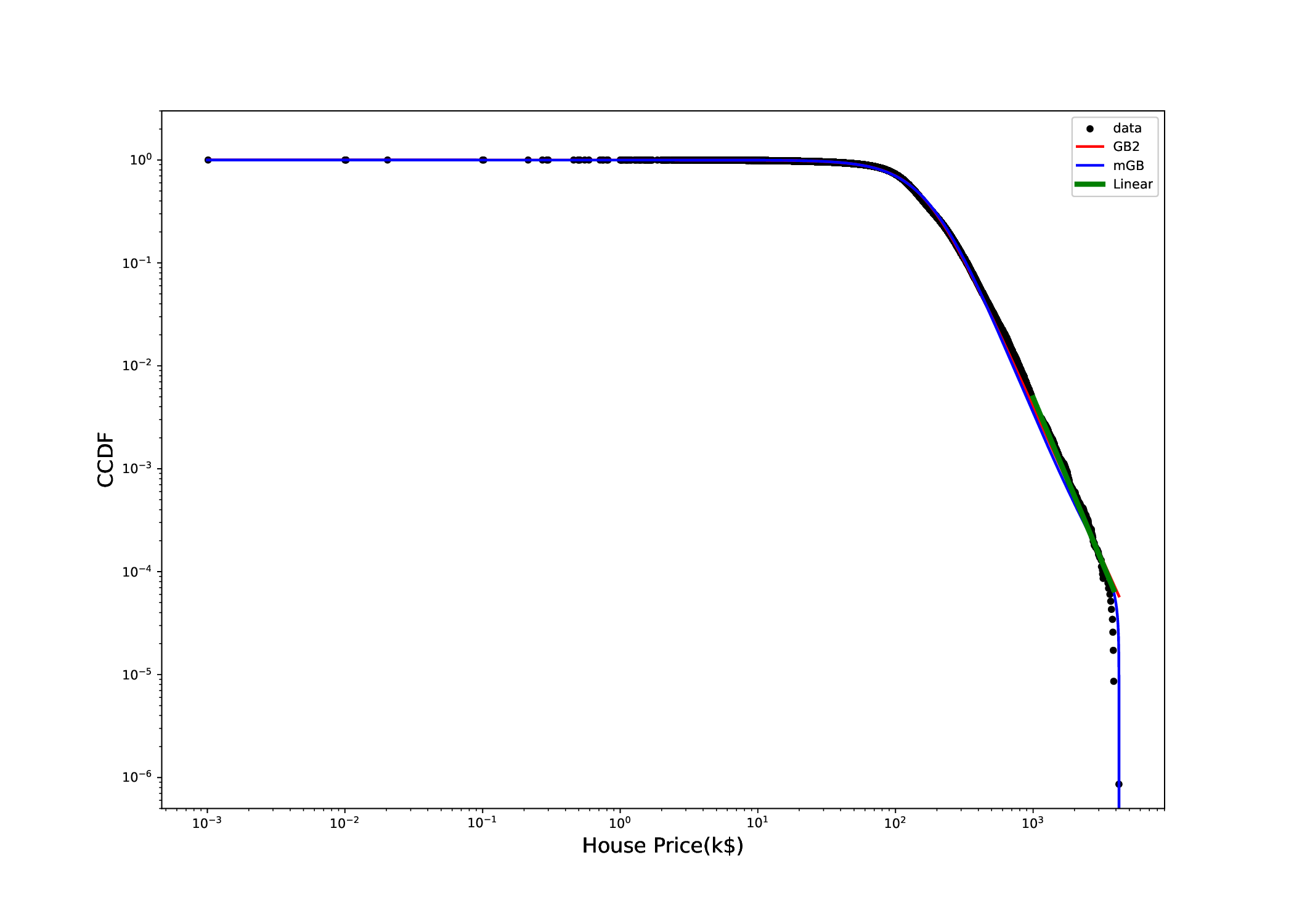}
		\includegraphics[width = .77 \textwidth]{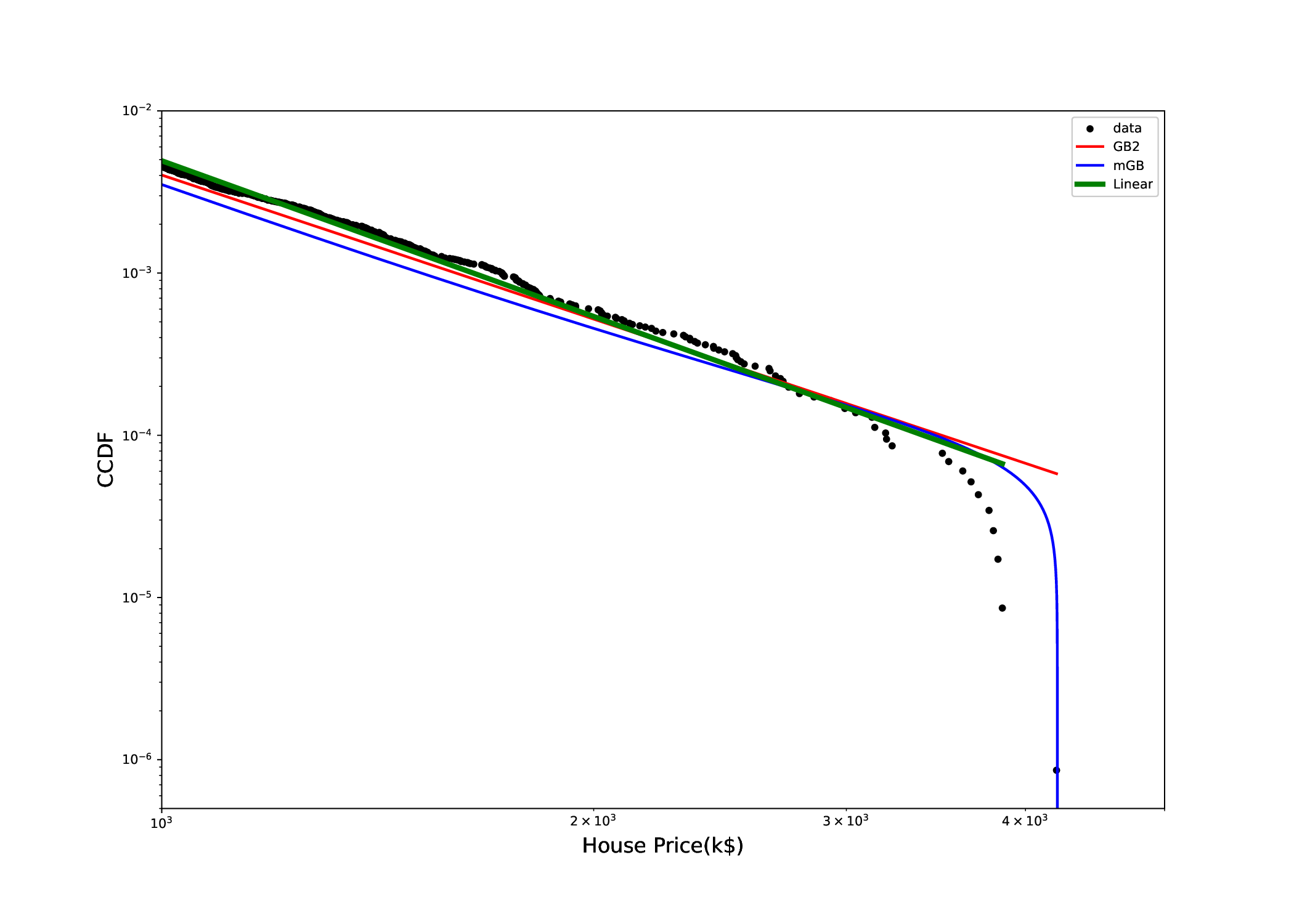}
	\caption{HP CCDF: mGB, GB2 fits of the full distribution and LF of the tail (top); tail area (bottom).}
	\label{CCDFHP}
\end{figure}

\newpage
Fig. \ref{CCDFHP} shows CCDF of the HP distribution on a log-log scale, with its mGB and GB2 fits and LF of the tail. The tail area is further expanded for a better view. Clearly the end points of the tail fall off abruptly from what looks like a rather well-defined power-law dependence, prompting a possibility that they may considered nDK.

\begin{figure}[htbp!]
	\centering
		\includegraphics[width = .77 \textwidth]{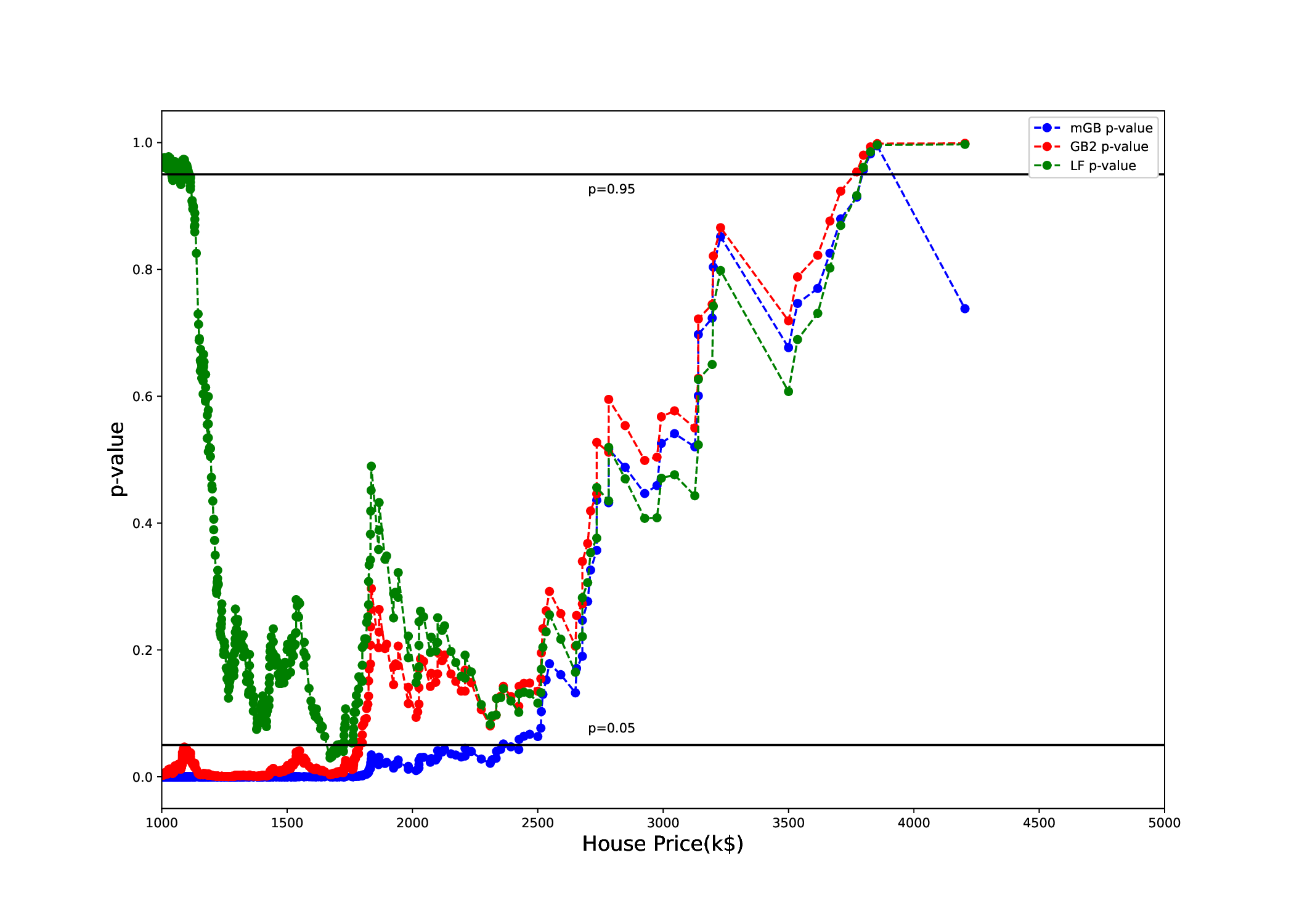}
		\includegraphics[width =.77 \textwidth]{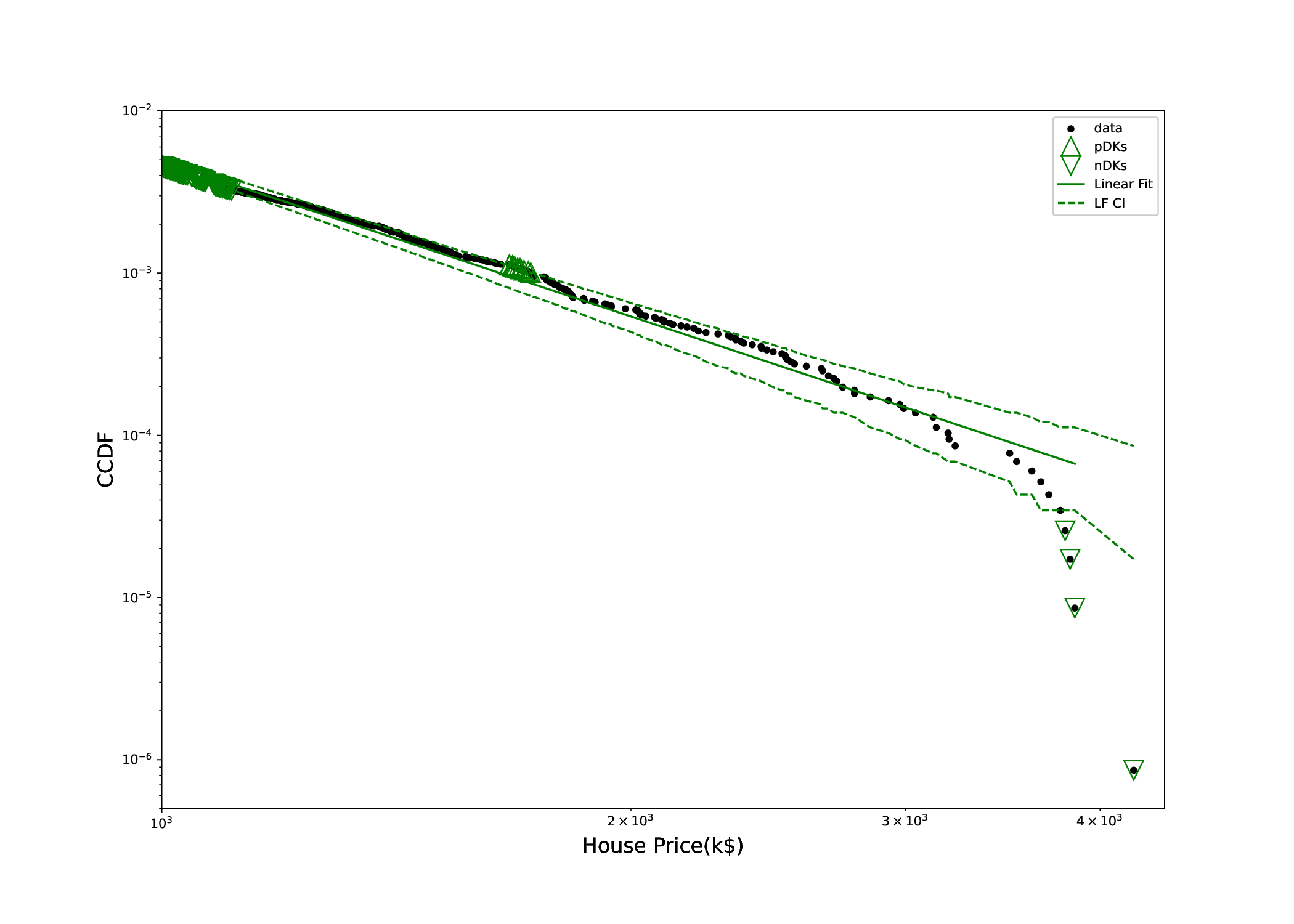}
	\caption{p-values from U test (top) and LF with CI (bottom): CI marked by dashed lines, pDK by up triangles, nDK by down triangles; LF excludes points visually deemed as possible nDK.}
	\label{pvalueHP}
\end{figure}

\newpage
Fig. \ref{pvalueHP} shows p-values obtained in U-test for all three fits (mGB, GB2 and LF), as well as the LF with its CI (dashed line). p-values such that $p<0.05$, had they been at the tail ends, could be considered DK, however in the earlier parts of the tail they are merely an indication of a poor fit (as are values $p>0.95$). We dubbed them "potential DK" (pDK) and mark in the plot with up triangles. p-values $p>0.95$ at the tail ends may qualify as actual nDK.

\begin{figure}[htbp!]
	\centering
		\includegraphics[width = .77 \textwidth]{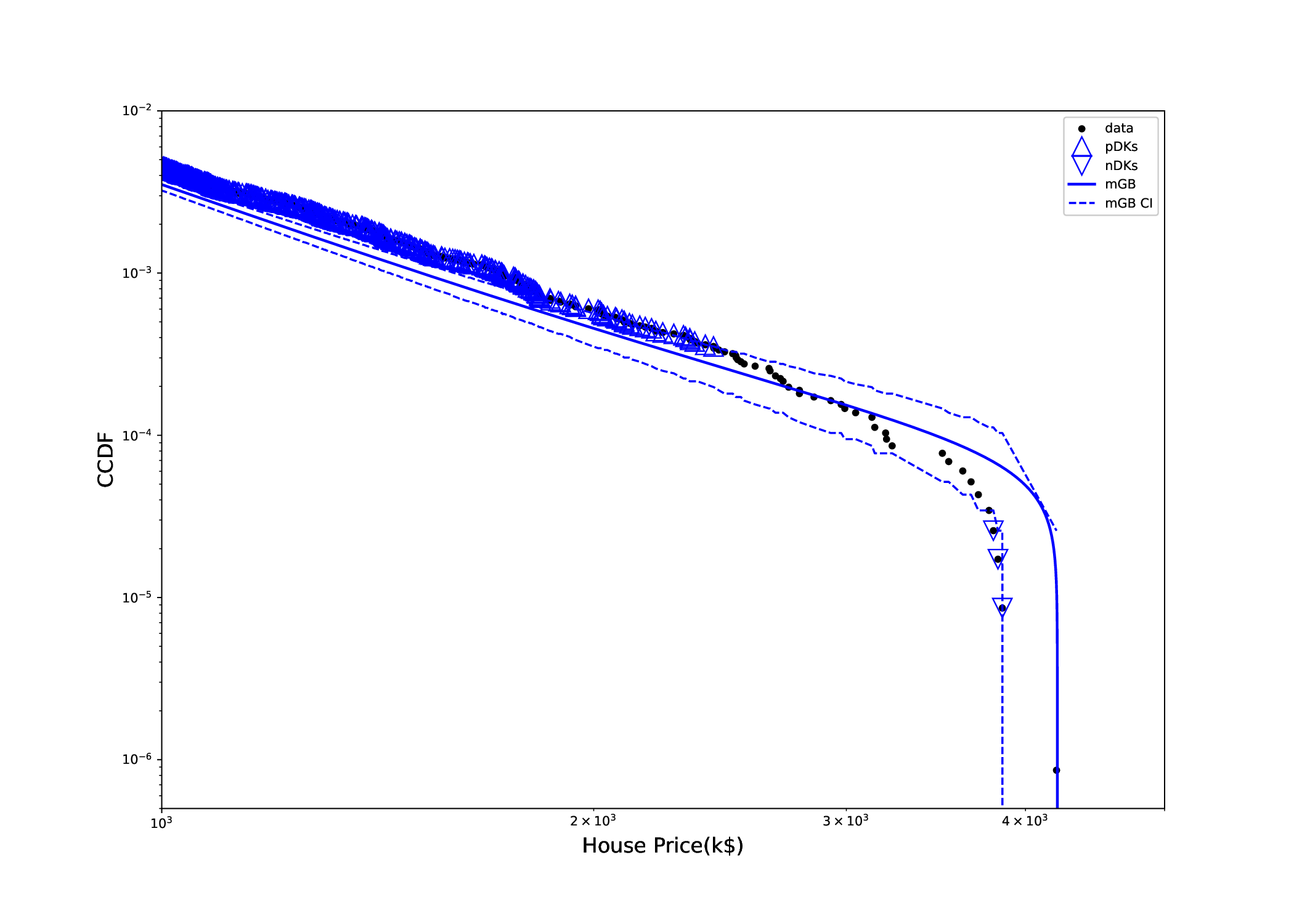}
		\includegraphics[width = .77 \textwidth]{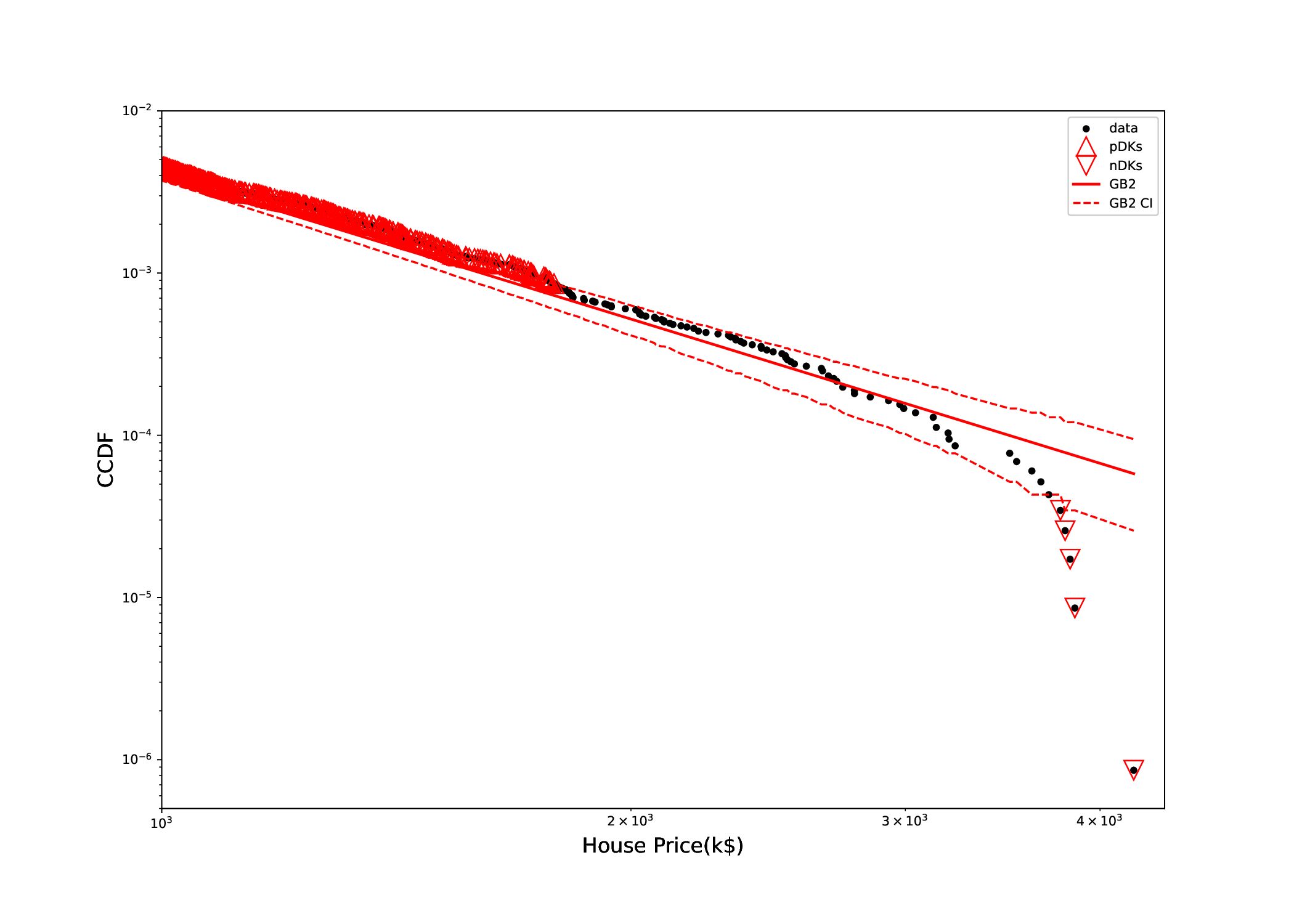}
	\caption{mGB fit with CI (top) and GB2 fit with CI (bottom) shown for tail area: CI marked by dashed lines, pDK by up triangles, nDK by down triangles.}
	\label{mGBGB2HP}
\end{figure}

\newpage
Fig. \ref{mGBGB2HP} shows tail areas of mGB and GB2 fits, with their CI, pDK, and nDK. Clearly mGB makes a decent attempt at approximating nDK behavior. However, earlier part of the tail (as well as full PDF above) indicate a rather poor fit by either.

\begin{figure}[htbp!]
	\centering
		\includegraphics[width = .77 \textwidth]{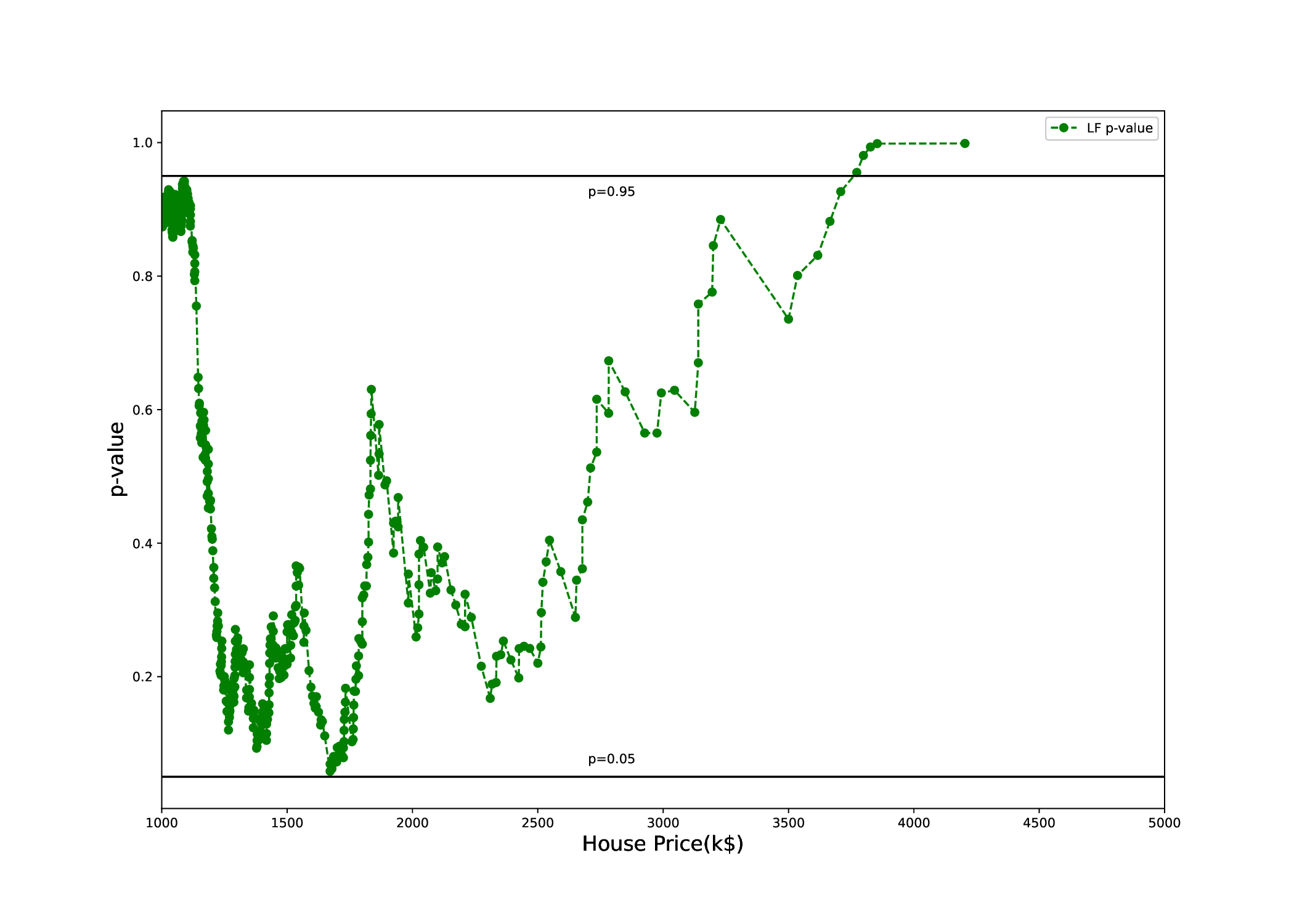}
		\includegraphics[width = .77 \textwidth]{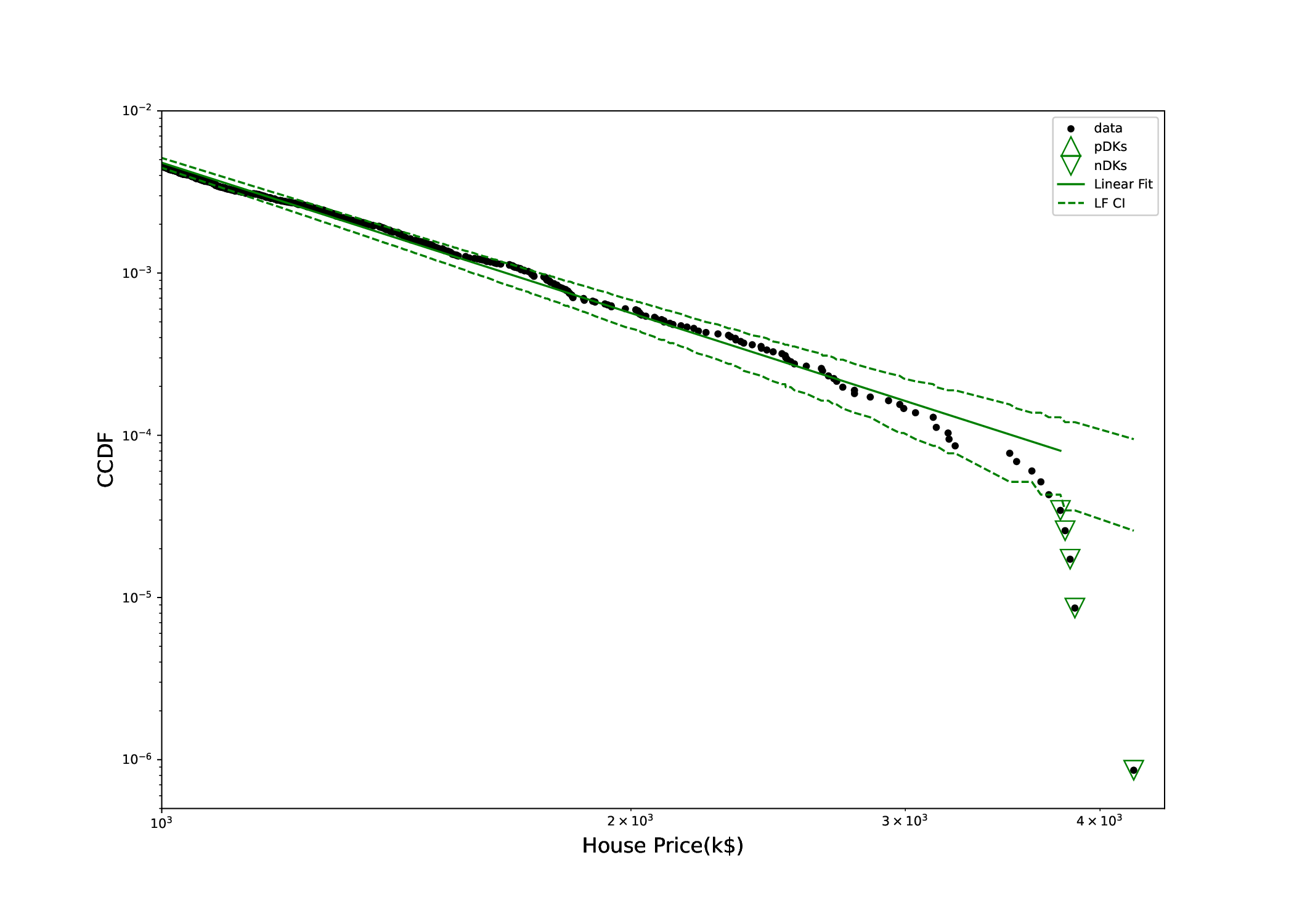}
	\caption{p-values from U test (top) and LF with CI (bottom): CI marked by dashed lines, pDK by up triangles, nDK by down triangles; LF excludes points with HP above 90\% of top HP.}
	\label{LF09HP}
\end{figure}

\newpage
Fig. \ref{LF09HP} shows p-value plot and LF, whose main difference from Fig. \ref{pvalueHP} is that LF is done by excluding the end points with HP greater than 90\% of the top price, rather than excluding possible outliers (nDK here) visually. Clearly, there is essentially very little difference between the two. It is the case for HPI as well, so we do not present a corresponding plot for HPI below.

\newpage
\subsection{Distribution of House Price Indicies \label{HPIfit}}
HPI are constructed using repeat-sale methodology \cite{baily1963regression}, \cite{case1987prices}, \cite{case1989efficiency}, \cite{calhoun1996house}, \cite{bogin2016local}. Specifically, we utilize FHFA HPI for nearly 18,000 US ZIP codes over a period of 40 years, starting in 1980's \cite{bogin2016local}. HPI are updated annually on FHFA site \cite{fhfa2003hpi}. HPI can be viewed as a proxy to HP, which in turn can be viewed as a proxy to income distribution -- here between different ZIP codes. In what follows, we initially analyze single-year distributions of HPI, followed by the multi-year distribution, which can be considered as a sum of single-year distributions.

\subsubsection{Single-Year Distribuions\label{2019}} 
We conducted fitting and analysis of a number of single-year HPI and found the results to be very similar, in particular over last two decades. Therefore here we present the results for only one year, 2019. The dataset contains HPI for 9033 zip codes. Table \ref{paramsHPI} contains all parameters of the fits, as well as slopes of GB2 and LF. %and KS statistic for mGB and GB2 fits; as mentioned above, the latter is meant to largely reflect on comparative goodness of fit between mGB and GB2.

\begin{table}[!htb]
\caption{All estimated parameters of mGB and GB2 fits of 2019 HPI distribution and slopes of GB2 and of LF}
\label{paramsHPI}
\centering
\begin{tabular}{c c c c c c c c c}
 Fit&$\alpha$ & $\beta_1$&$\beta_2$&p&q&Slope of CCDF\\%&KS&\\ 
\hline
mGB&2.2532 & 914.3328&54.0456& 71.5687&1.7586&&\\% 0.0144& \\   
\hline
GB2&1.6063 & & 58.029&  52.5168 & 5.6728&-9.1122&\\% 0.0169& \\  
\hline
LF& & && &&-7.1066&& \\  
%\hline
%LF-2& & && &&-6.9172&& \\ 
\end{tabular}
\end{table}

\begin{figure}[htbp!]
	\centering
		\includegraphics[width = .77 \textwidth]{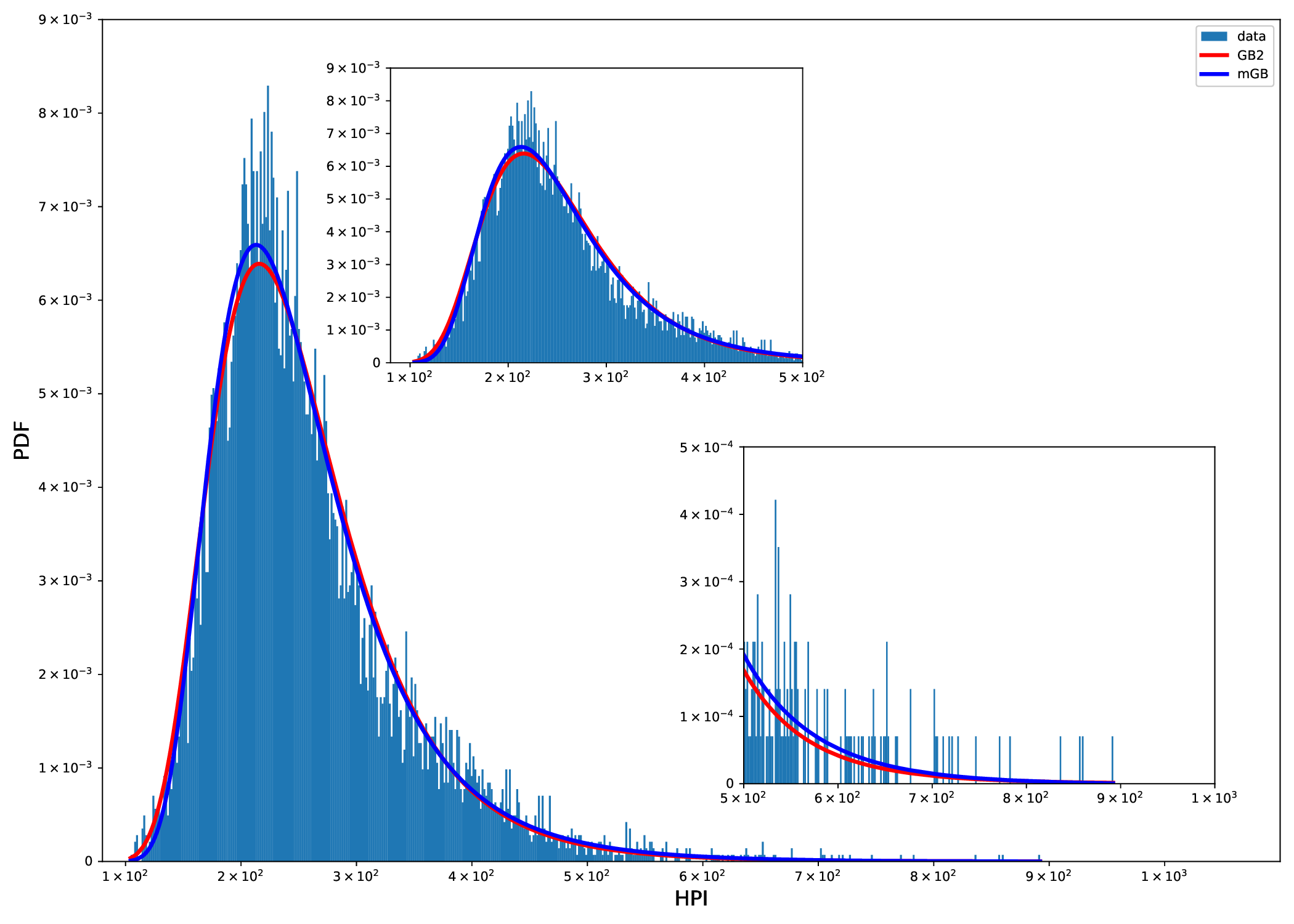}
	\caption{HPI 2019 PDF: mGB and GB2 fits.}
	\label{PDFHPI2019}
\end{figure}

Fig. \ref{PDFHPI2019} shows PDF of single-year HPI with mGB and GB2 fits. Visually these PDF fits are considerably better than those of HP.

\begin{figure}[htbp!]
	\centering
		\includegraphics[width = .77 \textwidth]{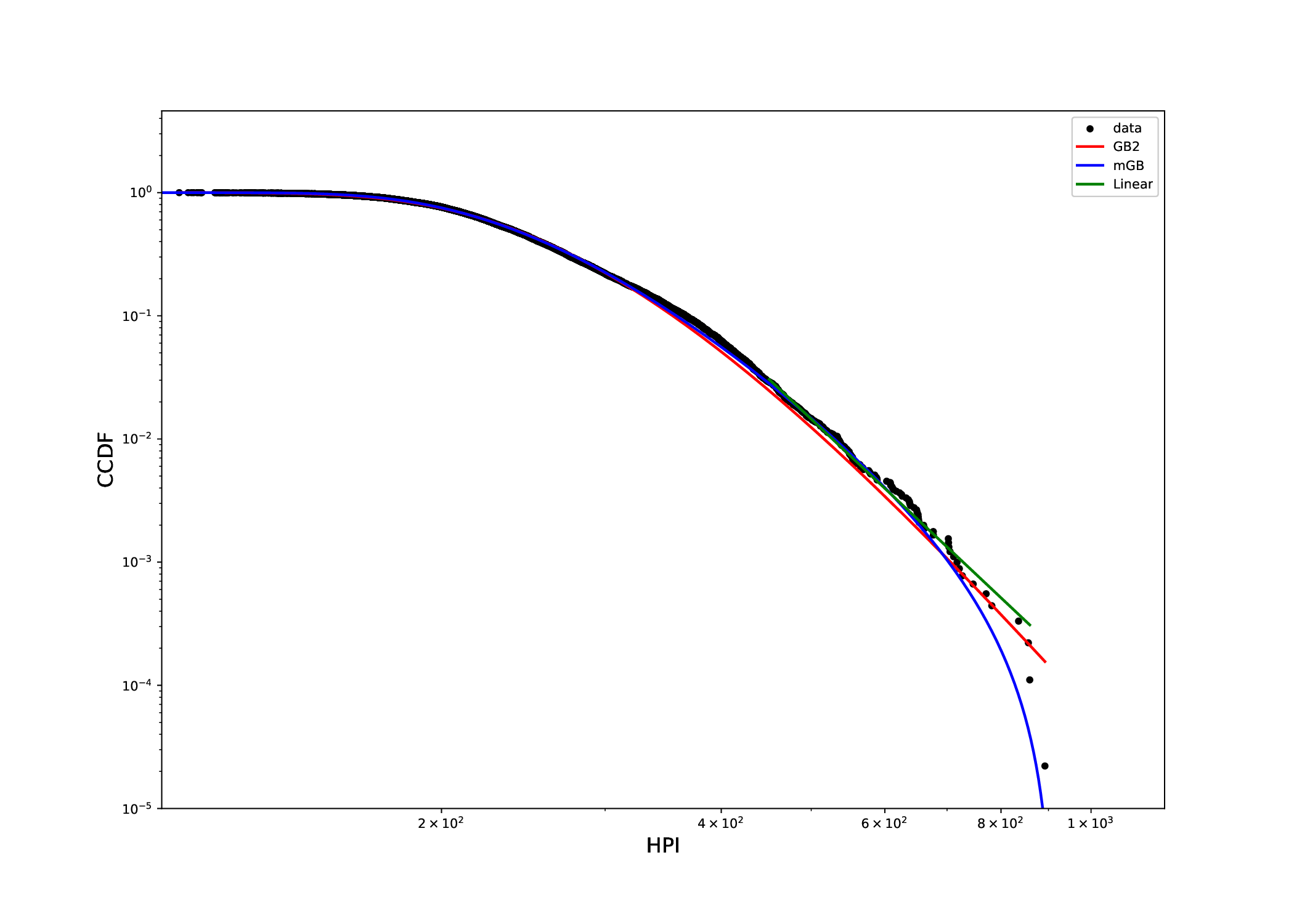}
		\includegraphics[width = .77 \textwidth]{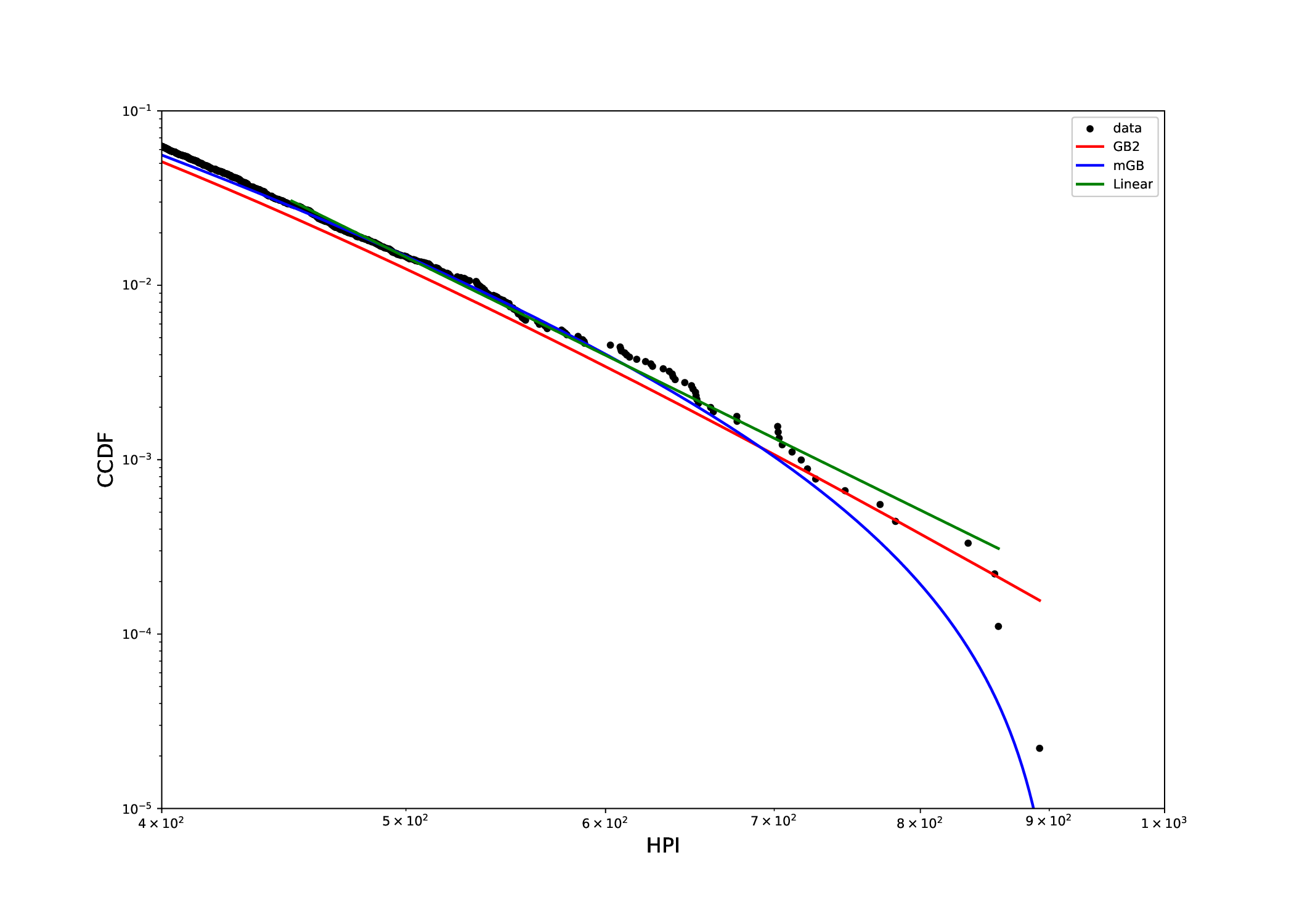}
	\caption{HPI 2019 CCDF: mGB, GB2 fits of the full distribution and LF of the tail (top); tail area (bottom). }
	\label{CCDFHPI2019}
\end{figure}

\newpage
Fig. \ref{CCDFHPI2019} shows CCDF of the HPI distribution on a log-log scale, with its mGB and GB2 fits and LF of the tail. The tail area is further expanded for a better view. Visually, the tail behavior is more consistent with power law than was the case of HP. This is further confirmed by p-values obtained in the U-test (see Fig. \ref{pvalueHPI2019} below).

\newpage
Fig. \ref{pvalueHPI2019} shows p-values obtained in U-test for all three fits (mGB, GB2 and LF), as well as the LF with its CI (dashed line). We observe that GB2 and LF do not have p-values $p>0.95$, which indicates less likelihood of outliers form the power-law behavior than was seen for HP, consistent with the previous visual observation in Fig. \ref{CCDFHPI2019}.

\begin{figure}[htbp!]
	\centering
		\includegraphics[width = .77 \textwidth]{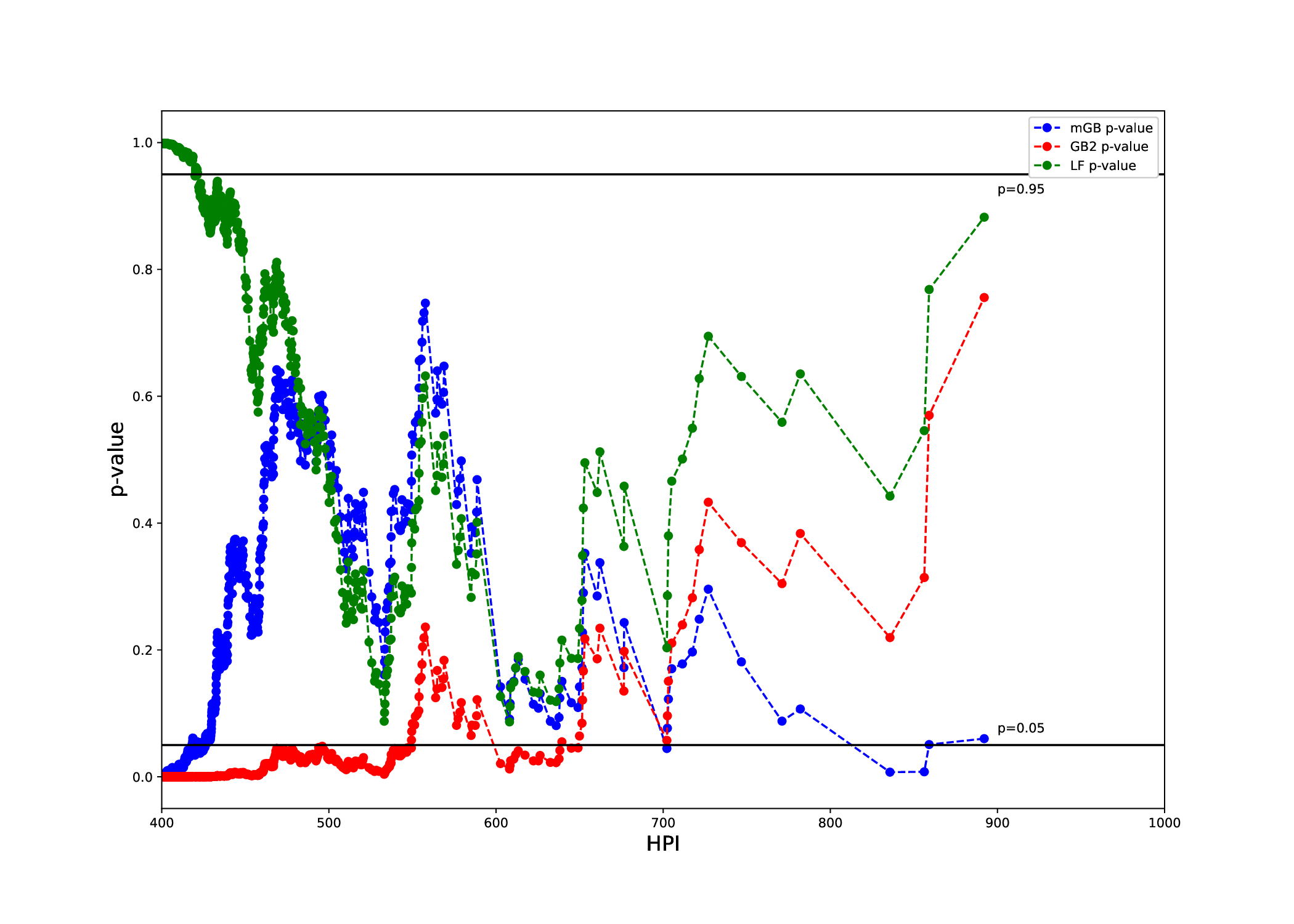}
		\includegraphics[width =.77 \textwidth]{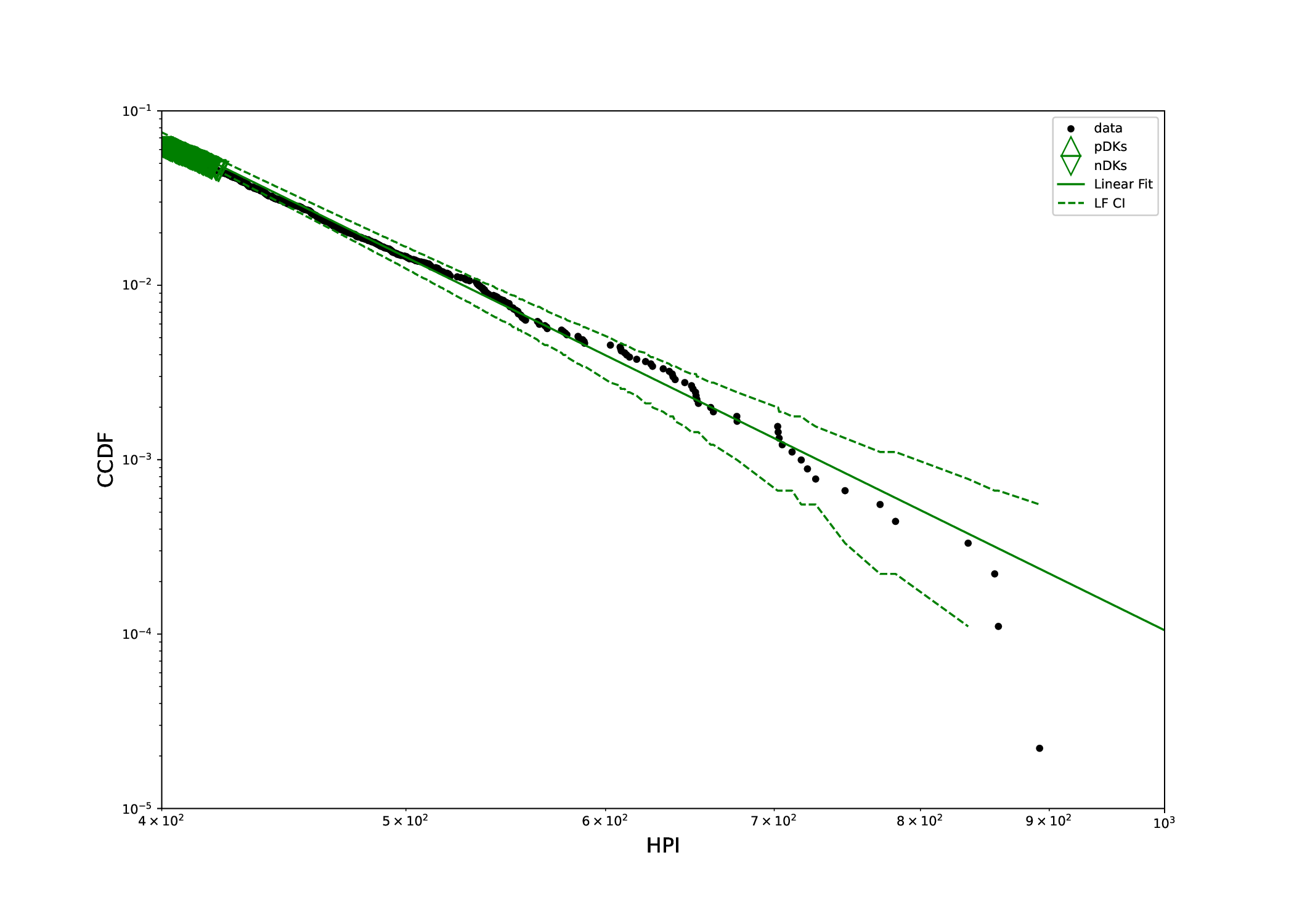}
	\caption{p-values from U test (top) and LF with CI (bottom): CI marked by dashed lines, pDK by up triangles, nDK by down triangles; LF excludes points visually deemed as possible nDK.}
	\label{pvalueHPI2019}
\end{figure}

\begin{figure}[htbp!]
	\centering
		\includegraphics[width = .77 \textwidth]{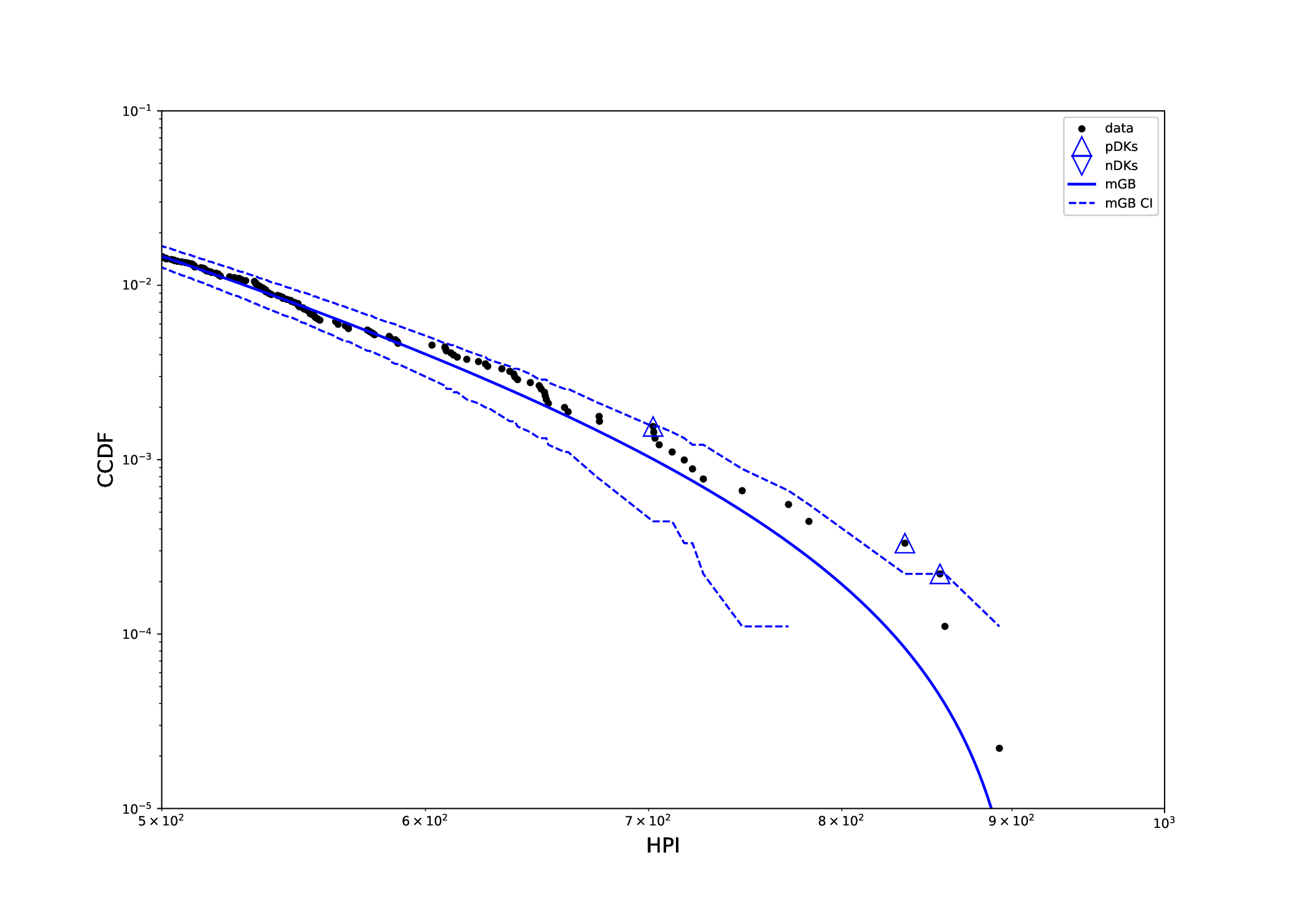}
		\includegraphics[width = .77 \textwidth]{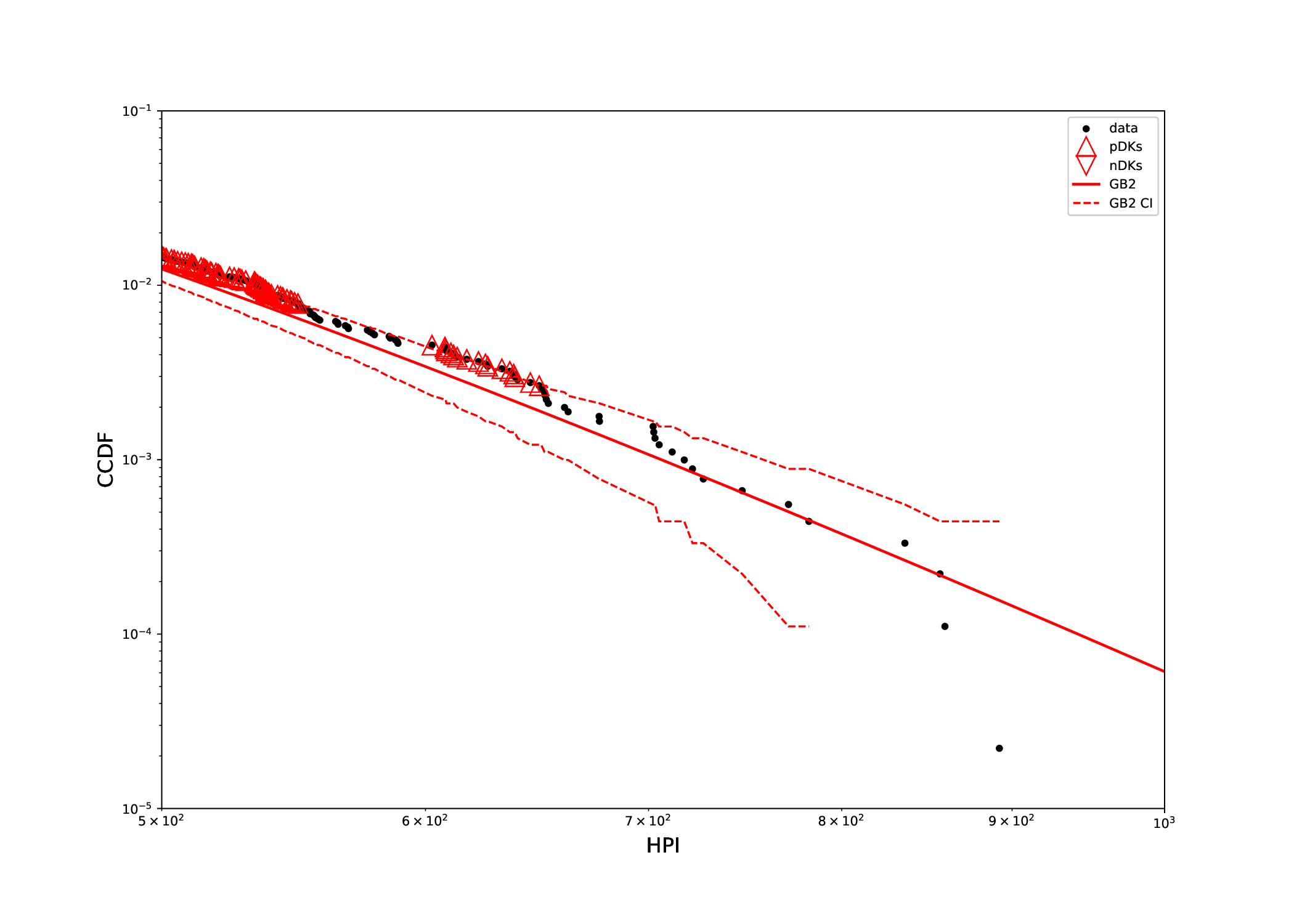}
	\caption{mGB fit with CI (top) and GB2 fit with CI (bottom) shown for tail area: CI marked by dashed lines, pDK by up triangles, nDK by down triangles.}
	\label{mGBGB2HPI2019}
\end{figure}

\newpage
Fig. \ref{mGBGB2HPI2019} shows tail areas of mGB and GB2 fits, with their CI, pDK, and nDK. Both distributions are relatively successful at approximating different parts of the tail.

\newpage
\subsubsection{Muli-Year Distibuions\label{2000+}} 
As was discussed in Sec. \ref{2019}, tails of single-year HPI align better with power law than those of HP. However, there is also an obvious trend towards moving down from the straight line at tails ends. Since single-year HPI also have an oder of magnitude fewer points than HP, we wanted to ascertain whether   this trend was significant. 

\begin{figure}[htbp!]
	\centering
		\includegraphics[width = .77 \textwidth]{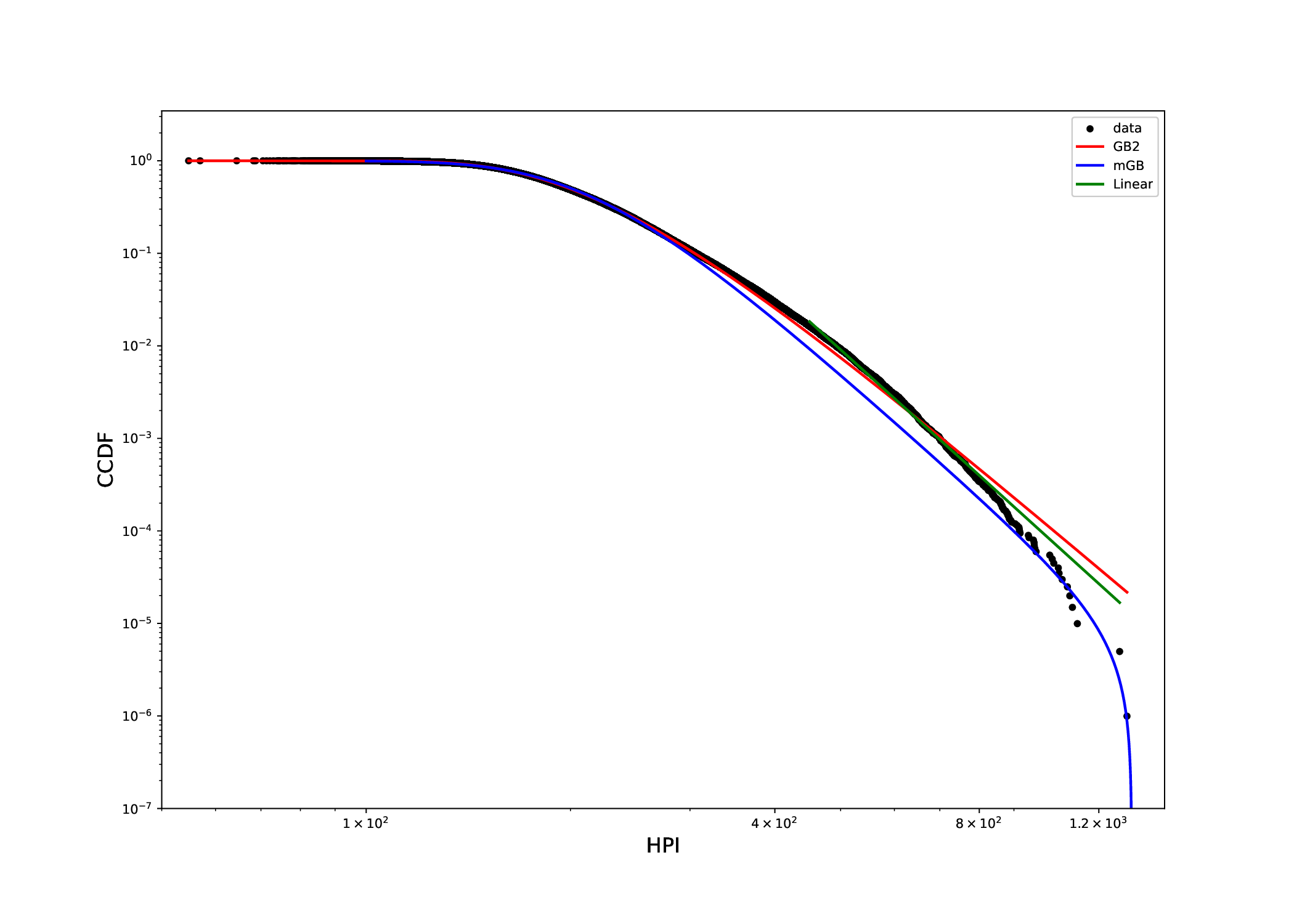}
		\includegraphics[width = .77 \textwidth]{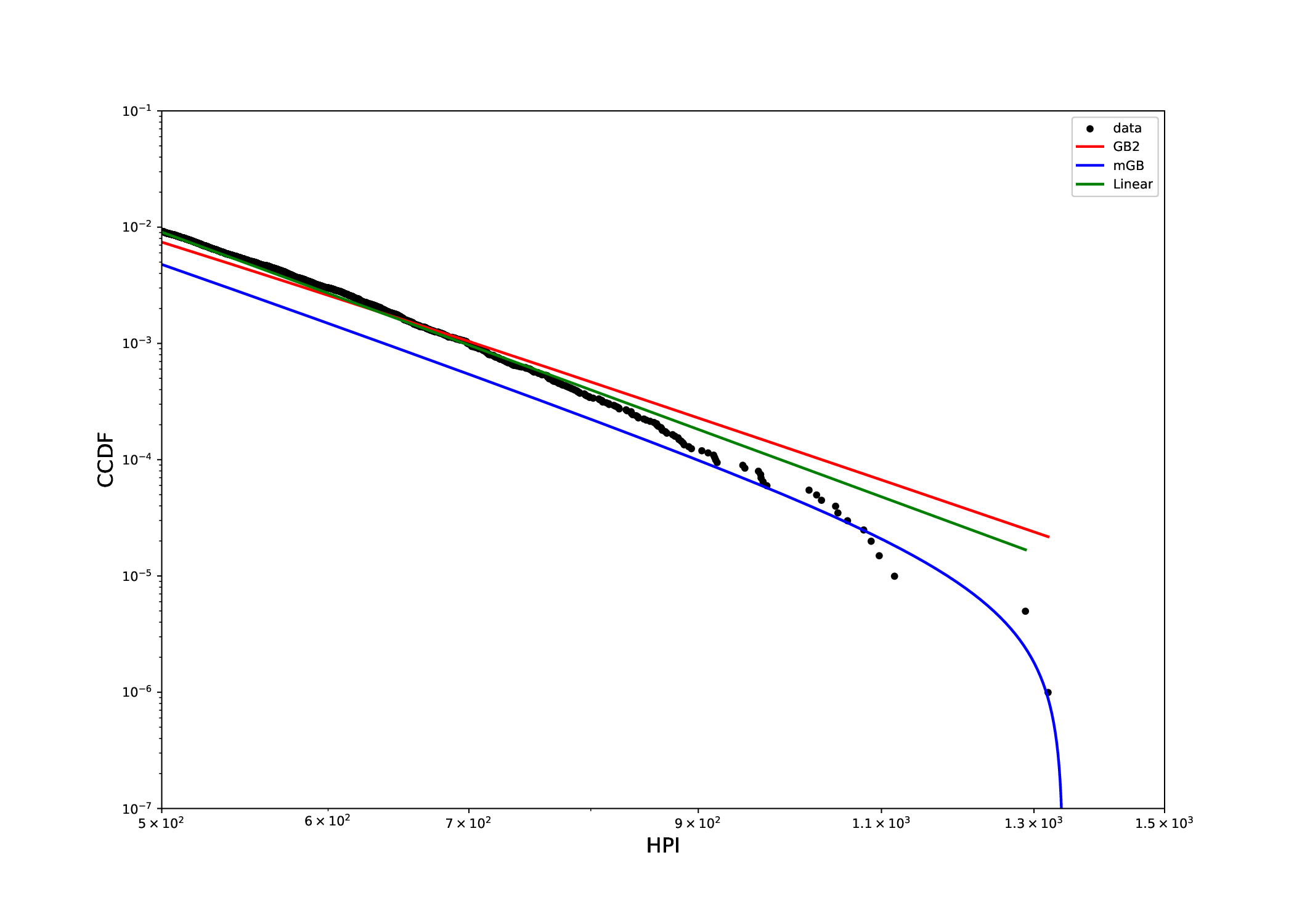}
	\caption{HPI 2000-2022 CCDF: mGB, GB2 fits of the full distribution and LF of the tail (top); tail area (bottom). }
	\label{CCDFHPI2000+}
\end{figure}

\begin{figure}[htbp!]
	\centering
		\includegraphics[width = .77 \textwidth]{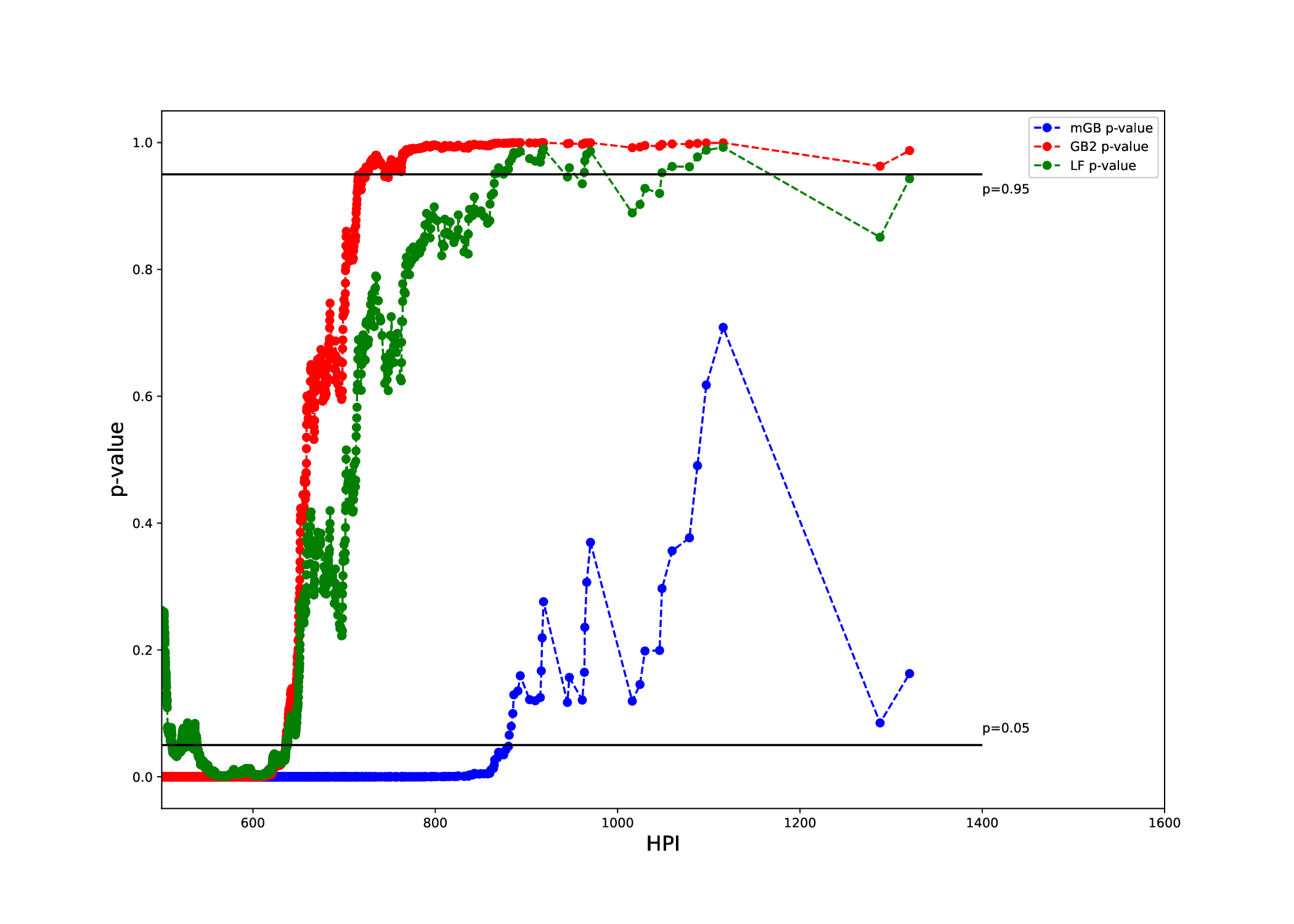}
		\includegraphics[width =.77 \textwidth]{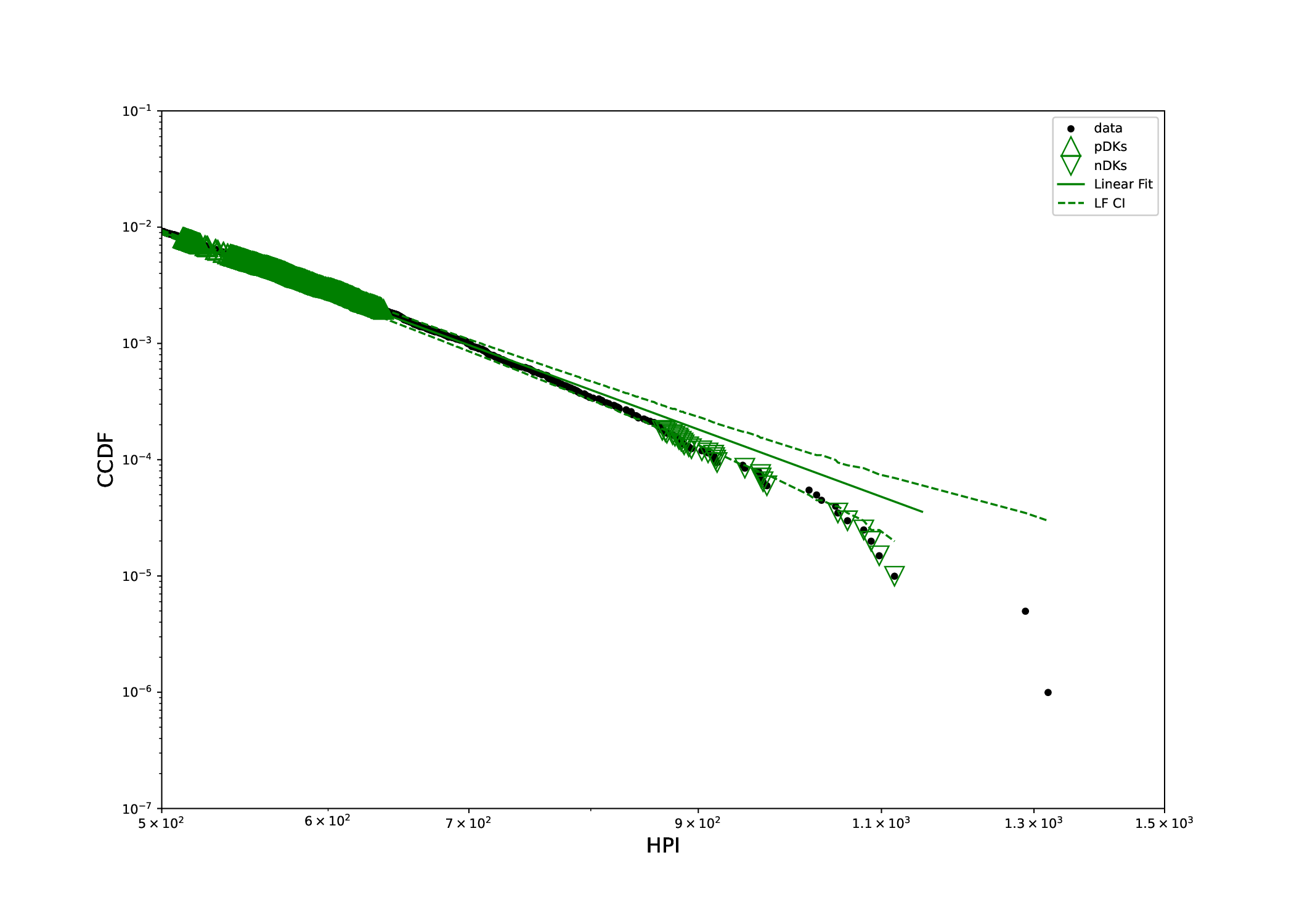}
	\caption{p-values from U test (top) and LF with CI (bottom): CI marked by dashed lines, pDK by up triangles, nDK by down triangles; LF excludes points visually deemed as possible nDK.}
		\label{pvalueHPI2000+}
\end{figure}

Towards this end we studied a combined multi-year distribution of HPI for years 2000-2022, which contained 201040 data points. The main result, as seen in Figs. \ref{CCDFHPI2000+} and \ref{pvalueHPI2000+} below is that the tails of the combined HPI is more aligned with the finite upper limit of HPI and, accordingly, with mGB distribution. Of course, such upper limit of the variable does not have to be fixed -- it may change as HPI is updated annually. 

Table \ref{paramsHPI+} contains all parameters of the fits of the multi-year distribution, as well as slopes of GB2 and LF.

\begin{table}[!htb]
\caption{All estimated parameters of mGB and GB2 fits of HPI distribution for years 2000-2022 and slopes of GB2 and of LF}
\label{paramsHPI+}
\centering
\begin{tabular}{c c c c c c c c c}
 Fit&$\alpha$ & $\beta_1$&$\beta_2$&p&q&Slope of CCDF&\\%KS&\\ 
\hline
mGB&3.3038 & 1342.3155&163.8916& 3.6162&1.0004&&\\%  0.0198& \\   
\hline
GB2&2.1786 & &  42.1151&  75.7226 &  2.8667&-6.2454&\\% 0.0073& \\  
\hline
LF& & && &&-6.6488&& \\  
%\hline
%LF-2& & && &&-6.6332&& \\ 
\end{tabular}
\end{table}

\begin{figure}[htbp!]
	\centering
		\includegraphics[width = .77 \textwidth]{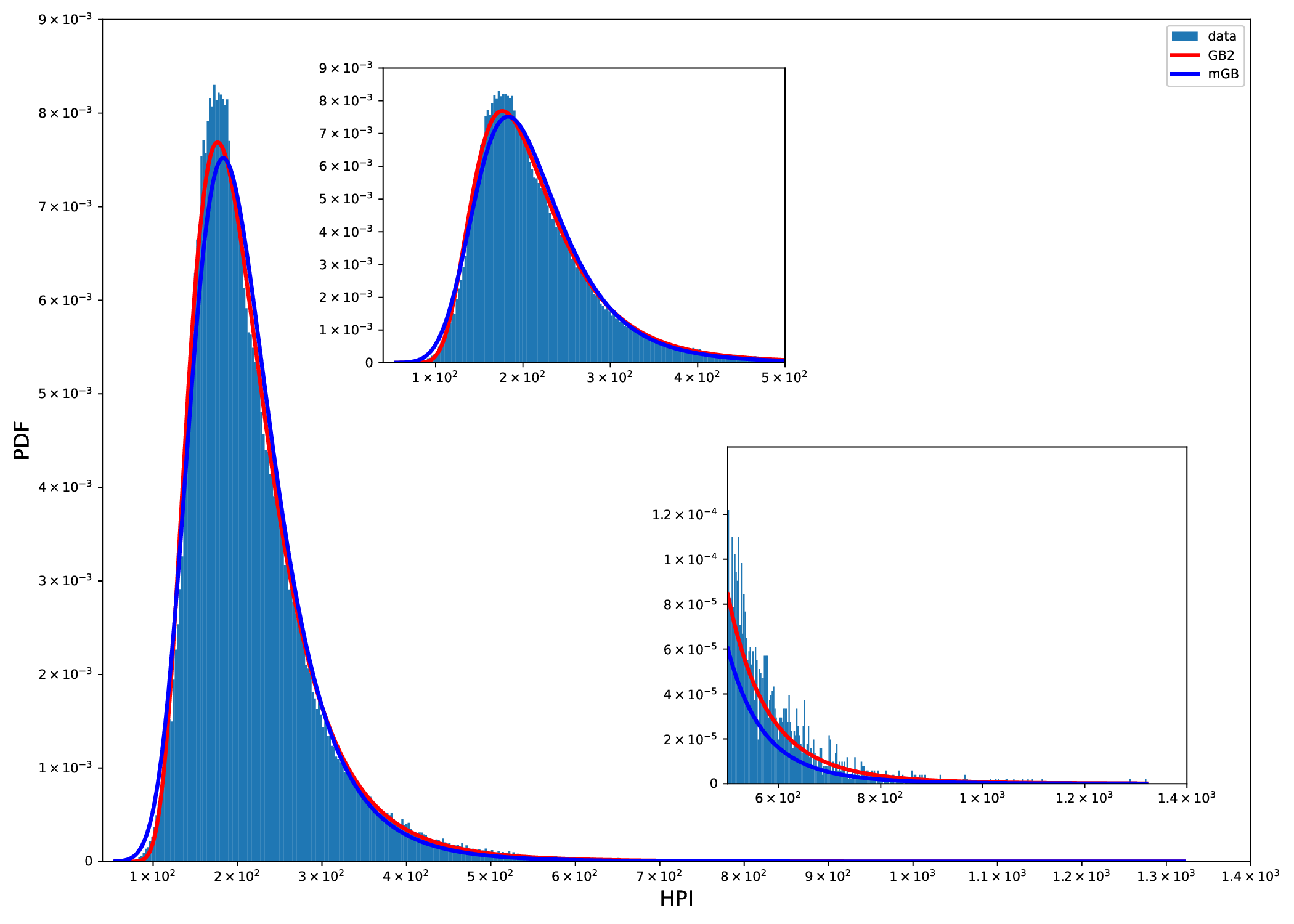}
	\caption{HPI 2000-2022 PDF: mGB and GB2 fits.}
	\label{PDFHPI2000+}
\end{figure}

Fig. \ref{PDFHPI2000+} shows PDF of multi-year HPI with mGB and GB2 fits. Visually the fits are better than their single-year counterpart, most likely because of a much larger dataset.

\newpage
Fig. \ref{CCDFHPI2000+} shows CCDF of the multi-year HPI distribution on a log-log scale, with its mGB and GB2 fits and LF of the tail. The tail area is further expanded for a better view. Visually, the tail behavior is more consistent with finite upper limit and mGB distribution than that of a single-year HPI, and is similar to HP. This is further confirmed by p-values obtained in the U-test (see Fig. \ref{pvalueHPI2000+} below).

\newpage
Fig. \ref{pvalueHPI2000+} shows p-values obtained in U-test for all three fits (mGB, GB2 and LF), as well as the LF with its CI (dashed line). We observe that, unlike GB2 and LF, mGB does not have p-values $p>0.95$ at tail ends, consistent with the previous visual observation in Fig. \ref{CCDFHPI2019}. Conversely, $p>0.95$ for GB2 and $p<0.05$ for mGB indicate poor fits at the onset of the tails.

\begin{figure}[htbp!]
	\centering
		\includegraphics[width = .77 \textwidth]{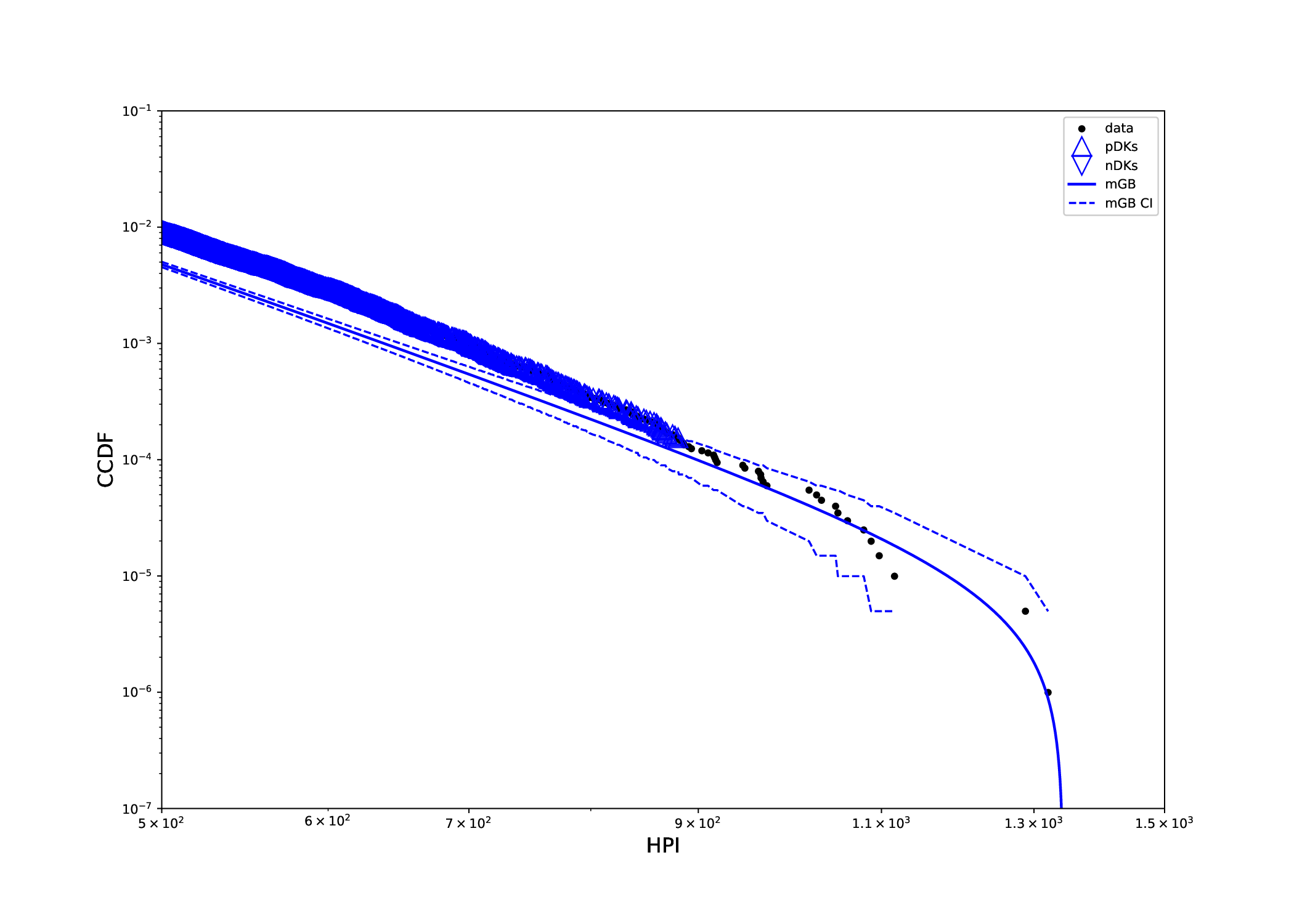}
		\includegraphics[width = .77 \textwidth]{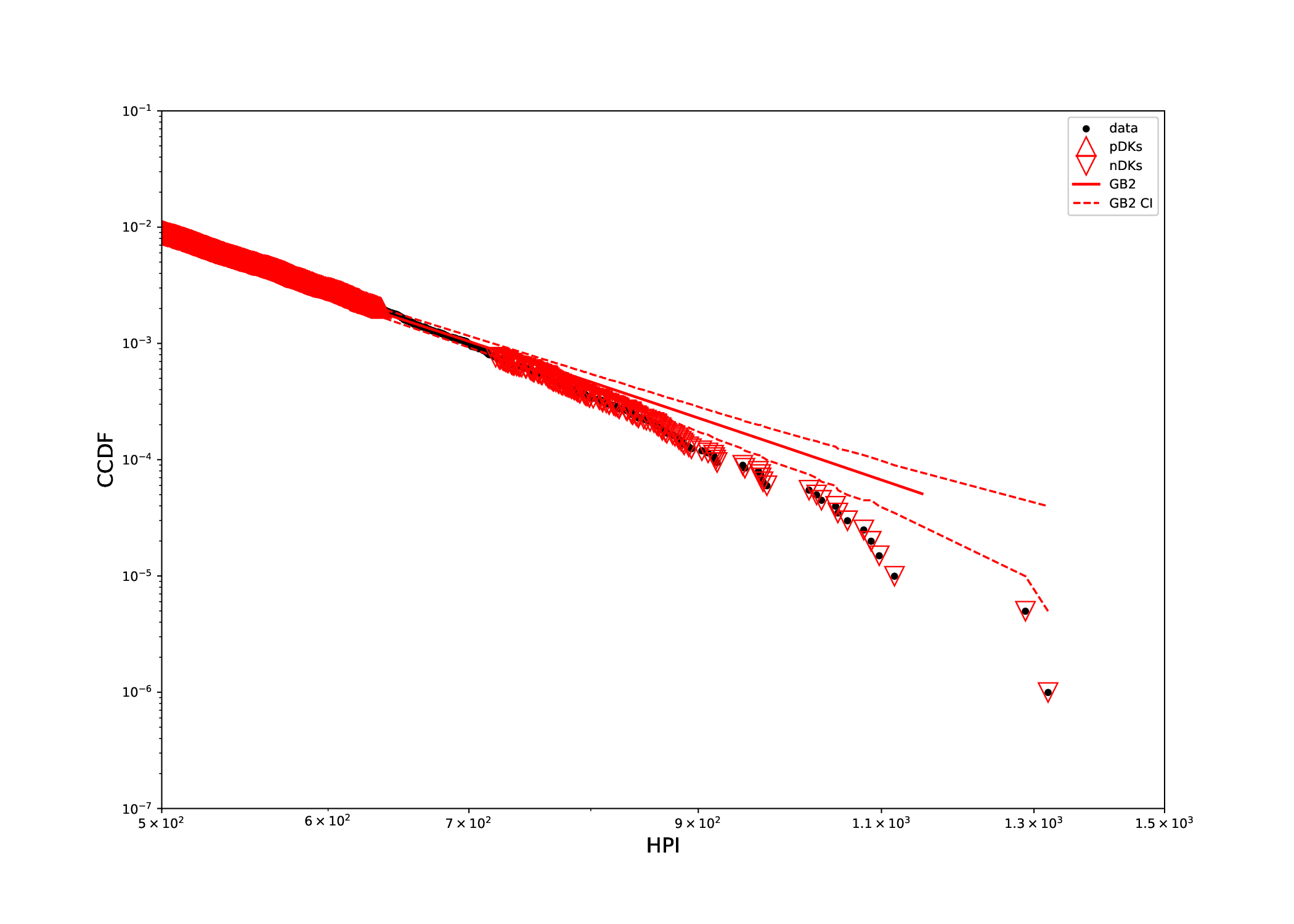}
	\caption{mGB fit with CI (top) and GB2 fit with CI (bottom) shown for tail area: CI marked by dashed lines, pDK by up triangles, nDK by down triangles.}
		\label{mGBGB2HPI2000+}
\end{figure}

\newpage
Fig. \ref{mGBGB2HPI2000+} shows tail areas of mGB and GB2 fits, with their CI, pDK, and nDK. Clearly mGB approximates rather well nDK behavior at the tail ends, which indicates a finite upper limit. However, as indicated above, the earlier part of the tail indicate a rather poor fit by either.

\newpage

\section{Discussion\label{Discussion}}

We studied distributions of house prices and house price indices. The key question we attempted to answer is whether those distributions have power-law (fat) tails or they are characterized by outliers at the tails ends, such as Dragon Kings and negative Dragon Kings. Towards this end, we conducted a linear fit of the tails of the complementary cumulative distribution functions on the log-log scale. In doing so, we excluded potential outliers - negative Dragon Kings here - and evaluated fits' confidence intervals, as well as conducted a U-test for the potential outliers which provides us with p-values that indicate whether they belong to the linear fit.

Based on the linear fit, confidence intervals, and p-values we can conclude that house price distributions and combined multi-year house price indices exhibit negative Dragon King behavior at the tail ends. Single-year house price indices, on the other hand, are not inconsistent with linear dependence, that is with power-law tails.

Our interest to house prices and house price indices was motivated by their being proxies to income distributions. Income distributions may be possible to describe by models of economic exchange, some of which can be reduced to stochastic differential equations with well-defined steady-state distributions. One class of such models results in steady-state distributions that belong to the Generalized Beta family of distributions. Therefore we also attempted to fit the entire empirical distributions with Generalized Beta Prime and modified Generalized Beta distributions: for a particular relationship between scale parameters the former is characterized by a power-law tail, while the latter follows the same power-law dependence, which is subsequently terminated at the finite value of the variable.

We find that modified Generalized Beta has some success in describing tail ends, however it is difficult to fully access overall goodness of fit. In particular we find that using Kolmogorov-Smirnov statistic for comparing goodness of fit between GB2 and mGB is rather inconclusive. Among idiosyncrasies are the following: despite an extra scale parameter, modified Generalized Beta may have large statistic than Generalized Beta Prime; for two close values of statistic the higher one may yield a better fit of the tail \footnote{ In this regard, for HPI we presented mGB and GB2 with slightly higher Kolmogorov-Statistic but with better tail fits.}; the largest values of statistic may occur prior to the onset of the tails. While there exist Monte-Carlo techniques aimed at estimating goodness of fit, we believe that they may be problematic, especially for distributions with power-law tails. We hope to address these issues in a future publication.

\section {Acknowledgments\label{Acknowledgments}}

We used Wolfram Mathematica in a subset of analytical calculations and MathWorks Matlab for much of the numerical work.

\section{Data Availability Statement\label{Data}}

Our dataset of Hamilton County, Ohio house prices is available upon request. Datasets of house price indices can be found at 

\url{https://www.fhfa.gov/DataTools/Downloads/Pages/House-Price-Index-Datasets.aspx#atvol}.
 
\section{Author Contribution Statement\label{Author}}

Jiong Liu and Hamed Farahani performed all numerical calculations, R.A. Serota was a lead on analytical part and on problem statement.

\section{Conflict of Interest Statement\label{Conflict}}
The authors have no conflicts of interest to declare.

\bibliography{mybib}

\begin{thebibliography}{10}
\expandafter\ifx\csname url\endcsname\relax
  \def\url#1{\texttt{#1}}\fi
\expandafter\ifx\csname urlprefix\endcsname\relax\def\urlprefix{URL }\fi
\expandafter\ifx\csname href\endcsname\relax
  \def\href#1#2{#2} \def\path#1{#1}\fi

\bibitem{chotikapanich2008modelling}
D.~Chotikapanich (Ed.), Modeling Income Distributions and Lorenz Curves,
  Springer, 2008.

\bibitem{mcdonald2008modelling}
J.~B. McDonald, Modeling Income Distributions and Lorenz Curves (Chotikapanich,
  Duangkamon - Editor), Springer, 2008, Ch. 3 and 8.

\bibitem{chotikapanich2018using}
D.~Chotikapanich, W.~E. Griffiths, G.~Hajargasht, W.~Karunarathne, P.~D.~S.
  Rao, Using the gb2 income distribution, Econometrics 6~(2) (2018) 21.

\bibitem{dashti2020stochastic}
M.~Dashti~Moghaddam, J.~Mills, R.~A. Serota, From a stochastic model of
  economic exchange to measures of inequality, Physica A 559~(125047) (2020).

\bibitem{sornette2012dragon}
D.~Sornette, G.~Ouillon, Dragon-kings: Mechanisms, statistical methods and
  empirical evidence, The European Physical Journal Special Topics 205 (2012)
  1--26.

\bibitem{wheatley2015multiple}
S.~Wheatley, D.~Sornette, \href{https://ssrn.com/abstract=2645709}{Multiple
  Outlier Detection in Samples with Exponential \& Pareto Tails: Redeeming the
  Inward Approach \& Detecting Dragon Kings} (2015).
\newline\urlprefix\url{https://ssrn.com/abstract=2645709}

\bibitem{liu2023rethinking}
J.~Liu, R.~A. Serota, Rethinking generalized beta family of distributions, The
  European Physical Journal B 96~(2) (2023) Aticle: 24.

\bibitem{liu2023dragon}
J.~Liu, R.~A. Serota, Are there dragon kings in the stock market?,
  arXiv:2307.03693 (2023).

\bibitem{baily1963regression}
M.~J. Bailey, R.~F. Muth, H.~O. Nourse, A regression method for real estate
  price index construction, Journal of American Statistical Association
  58~(304) (1963) 933--942.

\bibitem{case1987prices}
K.~E. Case, R.~J. Shiller, Prices of single family real estate prices, New
  England Economic Review (1987) 45--56.

\bibitem{case1989efficiency}
K.~E. Case, R.~J. Shiller, The efficiency of the market for single-family
  homes, The American Economic Review 79 (1989) 125--137.

\bibitem{calhoun1996house}
C.~A. Calhoun, Ofheo house price indexes: Hpi technical description, Working
  Paper N508, Office of Federal Housing Enterprise Oversight (1996).

\bibitem{bogin2016local}
A.~N. Bogin, W.~M. Doerne, W.~D. Larson, Local house price dynamics: New
  indices and stylized facts, Working Paper 16-01, Federal Housing Finance
  Agency,
  https://www.fhfa.gov/PolicyProgramsResearch/Research/Pages/wp1601.aspx
  (2016).

\bibitem{pisarenko2012robust}
V.~F. Pisarenko, D.~Sornette, Robust statistical tests of dragon-kings beyond
  power law distribution, The European Physical Journal Special Topics 205
  (2012) 95--115.

\bibitem{bouchaud2000wealth}
J.-P. Bouchaud, M.~M{\'e}zard, Wealth condensation in a simple model of
  economy, Physica A: Statistical Mechanics and its Applications 282~(3) (2000)
  536--545.

\bibitem{ma2013distribution}
T.~Ma, J.~G. Holden, R.~Serota, Distribution of wealth in a network model of
  the economy, Physica A: Statistical Mechanics and its Applications 392~(10)
  (2013) 2434--2441.

\bibitem{mcdonald1995generalization}
J.~B. McDonald, Y.~J. Xu, A generlazition of the beta distribution with
  applications, Journal of Econometrics 66 (1996) 133--152.

\bibitem{hertzler2003classical}
G.~Hertzler, "classical" probability distributions for stochastic dynamic
  models, in: 47th Annual Conference of the Australian Agricultural and
  Resource Economics Society, 2003.

\bibitem{dashti2021combined}
M.~Dashti~Moghaddam, R.~Serota, Combined mutiplicative-heston model for
  stochastic volatility, Physica A: Statistical Mechanics and its Applications
  561 (2021) 125263.

\bibitem{nist2022digital}
Nist digital library of mathematical functions, https://dlmf.nist.gov.

\bibitem{janczura2012black}
J.~Janczura, R.~Weron, Black swans or dragon-kings? a simple test for
  deviations from the power law⋆, European Physical Journal Special Topics
  205 (2012) 79--93.

\bibitem{fhfa2003hpi}
 [online]\href{https://www.fhfa.gov/DataTools/Downloads/Pages/House-Price-Index.aspx}{[link]}.

\end{thebibliography}

\end{document}